\documentclass[aps,prd,twocolumn,superscriptaddress,preprintnumbers,floatfix,nofootinbib,notitlepage,showkeys,showpacs]{revtex4-1}
\usepackage{multirow}
\usepackage{siunitx}
\usepackage{colortbl}

\usepackage{tikz}
\usetikzlibrary{decorations.markings, arrows, decorations.pathmorphing}

\usepackage{graphicx}
\usepackage{latexsym}
\usepackage{amsmath,amssymb}
\usepackage{amsfonts}
\usepackage{microtype}

\usepackage{mathtools}

\usepackage[export]{adjustbox}

\newlength{\FigureWidth}
\usepackage{xcolor}
\usepackage{makecell}
\newcommand\TopRule{\Xhline{0.08em}}
\newcommand\HeaderRule{\Xhline{0.05em}}
\newcommand\MidRule{\Xhline{0.03em}}
\newcommand\BotRule{\Xhline{0.08em}}

\newcommand\SetTableProperties{
  \renewcommand{\arraystretch}{1.4}
}

\usepackage{microtype}

\newcommand{\ua}{{\uparrow}}
\newcommand{\LX}{L_X}
\newcommand{\LT}{L_T}
\newcommand{\da}{{\downarrow}}

\usepackage{placeins}

\usepackage{hyperref}
\usepackage[capitalise]{cleveref}

\newcommand\<{\langle}
\renewcommand\>{\rangle}
\DeclareMathOperator{\Trace}{Tr}
\newcommand\MCConfig{\mathcal{C}}
\newcommand\Order{O}

\newcommand\mbar{\overline{m}}
\newcommand\EnergyFG{E_{\text{FG}}}
\newcommand\EnergyFermi{\epsilon_{\text{F}}}

\newcommand\EnergyFermionic{E^{\text{f}}}
\newcommand\EnergyNNOrderZero{E_{N_\ua, N_\da}^{(0)}}

\newcommand\EnergyGroundNNFermion{E_{N_\ua, N_\da}^{\text{f}\,0}}
\newcommand\EnergyGroundNNOrder[1]{E^{0, #1}_{N_\ua, N_\da}}
\newcommand\Hopping[1]{t_{#1}}
\renewcommand\vec\mathbf

\newcommand\NumHops[1]{{n_h^{#1}}}
\newcommand\NumParticles[1]{{N_{#1}}}
\newcommand\NumInteractions{{n_I}}

\newcommand\WeightMove{W_m}
\newcommand\WeightHop{W_h}
\newcommand\WeightInt{W_I}
\newcommand\NN{{N_\ua,N_\da}}

\usepackage{wormdiagrams}

\begin{document}
\title{Few-body physics on a space-time lattice in the worldline approach}

\author{Hersh Singh and Shailesh Chandrasekharan}
\address{Department of Physics, Box 90305, Duke University, Durham, NC 27708, USA}

\begin{abstract}
We formulate the physics of two species of non-relativistic hard-core bosons with attractive or repulsive delta function interactions on a space-time lattice in the worldline approach. We show that worm algorithms can efficiently sample the worldline configurations in any fixed particle-number sector if the chemical potential is tuned carefully. 
Since fermions can be treated as hard-core bosons up to a permutation sign, we apply this approach to study non-relativistic fermions. 
The fermion permutation sign is an observable in this approach and can be used to extract energies in each particle-number sector. 
In one dimension, non-relativistic fermions can only permute across boundaries, and so our approach does not suffer from sign problems in many cases, unlike the auxiliary field method. 
Using our approach, 
we discover limitations of the recently proposed complex Langevin calculations in one spatial dimension for some parameter regimes.  
In higher dimensions, our method suffers from the usual fermion sign problem. 
Here we provide evidence that it may be possible to alleviate this problem for few-body physics.
\end{abstract}

\maketitle

\section{Introduction}
\label{sec:intro}

Computing the properties of quantum systems containing fermions remains challenging especially when perturbative techniques begin to fail. Even in the case of few-body physics, where each particle is described by a large dimensional vector space, the free and the interacting parts of the Hamiltonian may be diagonalized by two very different basis vectors and the ground state in a given particle-number sector may be severely entangled in both these bases with no apparent small parameters. This problem is even more severe in quantum field theories, where these particles are created out of a vacuum that can itself be non-trivial, like in Quantum Chromodynamics (QCD). For this reason, studying the properties of low-mass hadrons remains a daunting challenge in lattice QCD \cite{HueWenLin2015}. 

In the context of QCD, an alternative approach has become exciting in recent years and is based on ideas of a low-energy effective field theory formulated using the symmetries of QCD and the fact that the vacuum breaks the chiral symmetry spontaneously \cite{Hammer:2018fbe,Machleidt20111}. This chiral effective field theory ($\chi$-EFT) is constructed using nucleons and pions as the low-energy degrees of freedom. 
The interactions are described by local operators constructed from the nucleon and pion fields, organized as a power series in the ratio of relevant energy scales, which acts as the small parameter.  
But even at leading orders in the effective field theory, computing the properties of low lying hadrons and nuclei within $\chi$-EFT can require non-perturbative calculations.  While analytic techniques based on resummations of Feynman diagrams are useful for up to three particles \cite{PhysRevLett.82.463}, numerical approaches especially based on Monte Carlo methods become necessary for higher number of particles \cite{PhysRevLett.120.122502,Lahde:2013uqa}. 

Non-perturbative state of the art Monte Carlo methods for few-body  problems are based on auxiliary field techniques and fixed node approximations.  Many variants of the algorithms have been developed over the years and for further details we refer the reader to recent reviews of the subject \cite{Lee:2008fa,Drut:2012md,Carlson:2014vla}.  While these auxiliary field quantum Monte Carlo (AFQMC) methods have a clear advantage in certain parameter regimes, they also exhibit several limitations in other regimes \cite{Lahde:2015ona}. Difficulties of the method become apparent in one spatial dimension, which has become experimentally interesting in recent years, thanks to our ability to design and control ultracold quantum gases confined to optical traps \cite{Gross995,Bloch2005}. Also, many interesting quantum phenomena in higher dimensions have analogues in one spatial dimension \cite{PhysRevLett.109.250403,PhysRevLett.120.243002,PhysRevA.82.043606,PhysRevA.87.063617,PhysRevA.97.043630}. Motivated by this, the AFQMC method was recently used to study two species of fermions in one spatial dimension interacting through a delta function interaction \cite{PhysRevA.96.033635,PhysRevA.92.013631,PhysRevA.92.063609}.

One major limitation of the auxiliary field approach is that it suffers from sign problems in the presence of repulsive interactions or mass- and spin-imbalanced systems away from half filling. This is true even in one spatial dimension. 
In order to explore a solution to this problem, the auxiliary field technique was combined with the complex Langevin (CL) method to overcome the sign problem in Ref.~\cite{PhysRevD.96.094506}. Recently, this approach was also extended to higher dimensions \cite{PhysRevLett.121.173001}. Unfortunately, it is well known that CL methods may have uncontrolled systematic errors and can converge to wrong results \cite{PhysRevD.81.054508}. In fact, the authors of Ref.~\cite{PhysRevD.96.094506} suggested caution for their results, especially in the repulsive case where the CL approach showed fat-tailed distributions.  However, the authors wondered if the flattening of the ground-state energy as a function of the strength of repulsion, observed using the CL approach, was a sign of some interesting non-perturbative physics. The only way to be sure is to compare the results with other reliable methods. 

One spatial dimension is an excellent place to test methods like CL, since entanglement is greatly reduced and a variety of other methods that do not suffer from sign problems are usually available. For example, exact analytic calculations based on the Bethe ansatz are possible in some special cases \cite{yang_exact_1967, gaudin_systeme_1967, lieb_absence_1968, Turb2016}. With open boundary conditions or odd number of fermions with periodic boundary conditions, sign problems are absent in the worldline formulation \cite{wiese_bosonization_1993,evertz_loop_2003}. 
Recently the two-dimensional lattice Thirring model with open boundary conditions was formulated using the worldline method \cite{PhysRevD.97.054501} to provide benchmark results to test the Lefschetz thimble approach \cite{PhysRevD.95.014502}.
Thus, we should also be able to test the recent CL results of Ref.~\cite{PhysRevD.96.094506} using a similar worldline approach.

Motivated by this, we construct a general worldline approach to study quantum mechanics of hard-core bosons in any dimension. We show that we can use worm algorithms to update the worldline configurations efficiently in any particle-number sector by tuning the chemical potential carefully \cite{PhysRevLett.87.160601}. While worldline methods to study bosonic quantum field theories are well known by now \cite{Chandrasekharan:2008gp,Wolff:2010zu,Gattringer:2014nxa}, and have been used in several studies so far \cite{PhysRevD.81.125007,PhysRevD.94.114503,PhysRevLett.113.152002,Gattringer2018,Bruckmann:2015sua,PhysRevLett.120.241601}, their applicability to bosonic quantum many-body physics has remained relatively unexplored \cite{Batr2009,PhysRevA.85.063624}. Thus, our work can be viewed as an attempt to fill this gap.
Our method can easily be extended to fermions if we can compute the fermion permutation sign accurately. In one spatial dimension, fermions are identical to hard-core bosons up to boundary effects. Hence, our method can be used to check the recent results of Ref.~\cite{PhysRevD.96.094506}.  We find that the CL method yields incorrect results as the repulsive coupling strength grows, implying that the observed flattening is unphysical and an artifact of the method. 

We can easily adapt our approach to study fermionic particles even in higher dimensions by treating the fermionic permutation sign as an observable, but it becomes difficult to compute it accurately. As expected, this observable suffers from a severe signal-to-noise ratio problem, especially at low temperatures and when the number of particles becomes large. However, we provide evidence that at intermediate temperatures and with a small number of particles ($N \sim 10$) we may be able to beat the signal-to-noise ratio. This may allow us to explore new and interesting questions in few-body physics that are difficult to answer with the AFQMC method.

Our paper is organized as follows. In \cref{sec:lattice-model} we discuss the lattice Hamiltonian formulation of two species of hard-core bosons with a contact interaction. In \cref{sec:worldline-formulation}, we construct the worldline formulation of the problem on a space-time lattice and provide details of our worm algorithm in \cref{sec:worm-algorithm}. In \cref{sec:sign}, we discuss how we can study fermions by measuring the fermionic permutation sign. We also provide evidence that the fermion sign problem is mild in three dimensions for up to ten particles at an intermediate temperature. In \cref{sec:1d-results}, we present our results for fermions in one spatial dimension and consider two cases: the mass-balanced case and the mass-imbalanced case. In the mass-balanced case, we show that our results are in agreement with the exact results obtained using the Bethe ansatz. We also show that in both cases the flattening of the ground state-energy observed in the CL calculations is absent in our approach. In \cref{sec:extractGSE}, we discuss a limitation of the traditional approach used to extract the ground state energy and suggest a complementary method that is more reliable. In \cref{sec:higher-dimensions}, we provide evidence that we can also compute fermionic ground state energies in $3+1$ dimensions by reproducing a benchmark calculation performed several years ago in Ref.~\cite{PhysRevA.83.063619} and present our conclusions in \cref{sec:conclusions}.

\section{Lattice Model}
\label{sec:lattice-model}

The physics of two species of non-relativistic hard-core bosons that we consider in this work can be represented through the lattice Hamiltonian
\begin{align}
    H &= -\sum_{i,\hat\alpha,\sigma} \ \Hopping{\sigma}\ (c^\dagger_{i,\sigma} c_{i+\hat{\alpha},\sigma}^{\phantom{dagger}} + c^\dagger_{i+\hat{\alpha},\sigma} c_{i,\sigma}^{\phantom{\dagger}} - 2
c^\dagger_{i,\sigma} c_{i,\sigma}^{\phantom{\dagger}})  
      \nonumber \\
    &\quad + \frac{U}{a^d} \ \sum_i \ \NumParticles{i,\ua} \NumParticles{i,\da},
\label{eq:lattice-model}
\end{align}
where $c^\dagger_{i,\sigma}, c_{i,\sigma}^{\phantom{\dagger}}$ create and annihilate bosons 
of species, say, spin-up ($\sigma=\ua$) or spin-down ($\sigma=\da$), 
at the site $i$ on a $d$-dimensional spatial lattice and 
$\hat{\alpha}$
represents unit vectors in the $d$ spatial directions. 
The site occupation-number operators are defined as $N_{i,\sigma} = c^\dagger_{i,\sigma} c_{i,\sigma}^{\phantom{\dagger}}$. 
The parameters $\Hopping{\sigma} = 1/(2 m_\sigma a^2)$ give the hopping strength of the particles 
in terms of the masses of the particles $m_\sigma$ and the lattice spacing $a$. 
The parameter $U$ is the bare interaction strength.
In addition, we impose the hard-core boson constraint on the states of the Hilbert space, which means each site can either be empty or contain a single boson of a particular species. In this work, we will study finite spatial boxes of size $\LX$ with periodic boundary conditions, so that the physical box size is given by $L = \LX a$. 
If the particles are taken to be fermions instead of hard-core bosons, the lattice model \eqref{eq:lattice-model} is known as the Hubbard model.

The lattice model \eqref{eq:lattice-model} has a naive continuum limit as $a\rightarrow 0$. In $d=1$, this limiting procedure leads to the continuum theory with a delta function interaction
\begin{align}
  H_{a \to 0} &= - \sum_{\sigma=\uparrow,  \downarrow} \frac{1}{2m_\sigma} \int\!\! dx \ \psi^\dagger_\sigma(x) \Big(\frac{d^2}{dx^2}\Big)\psi_\sigma(x) \nonumber \\
             &\quad\quad + g \int\!\! dx \ \psi^\dagger_\ua(x) \psi_\ua(x)\ \psi^\dagger_\da(x)\psi_\da(x)
               \label{cmodel}
\end{align}
where $g=U$. However, in higher dimensions $(d>1)$ the problem of the continuum limit is more subtle and the framework of effective field theory becomes necessary to implement it. For example, with equal-mass fermions in three dimensions, the coupling $U(a)$ can be tuned to the unitary fixed point to get an interacting field theory in the continuum limit. 
With bosons or additional quantum numbers, we can get Efimov physics which necessitates additional counterterms to renormalize the theory \cite{PhysRevLett.82.463}. 
For this reason, we confine the discussion of the continuum limit to one dimension. 
In higher dimensions we will simply view \eqref{eq:lattice-model} as a lattice model and set $a=1$.

In this work, we will compute the ground-state energy $E^0_{N_\ua,N_\da}$, which is the lowest eigenvalue of the Hamiltonian in \cref{eq:lattice-model} in the subspace with particle numbers $N_\ua$ and $N_\da$. One way to accomplish this is to simply compute the average energy
\begin{align}
    \langle E \rangle = \frac{1}{Z_\mu}\ \Trace \Big( \ H\ {e}^{-\beta H_\mu }\Big)
    \label{eq:avgE}
\end{align}
at very low temperature ($\beta \to \infty$), where we use the modified lattice Hamiltonian,
\begin{align}
    H_\mu = H - \sum_{i,\sigma} \mu _\sigma \NumParticles{\sigma,i},
\end{align}
with chemical potentials $\mu_\sigma$ for the two particle species, and the partition function 
\begin{align}
    Z_\mu = \Trace\Big( e^{-\beta H_\mu }\Big)
    \label{eq:Z-mu}
\end{align}
to define the expectation value. If we compute the trace in a fixed particle-number subspace, the chemical potential terms drop out and we indeed get $\langle E \rangle = E^0_{N_\ua,N_\da}$ in the limit $\beta \rightarrow \infty$. Thus, we need a method to compute $\langle E\rangle$ reliably for large values of $\beta$.

In the next two sections, we will find an expression for $\langle E\rangle$ in the worldline formulation and construct a worm algorithm to compute it efficiently. Worm algorithms work by adding and removing particles which then updates the worldline configuration. If energetics do not favor this, algorithms can develop exponentially long autocorrelation times. 
Hence, for efficient sampling in a fixed particle-number sector it is important to tune the chemical potentials carefully. 
To understand why this is the case, let us consider the energy of the ground state containing $N_\ua$ and $N_\da$ particles in the presence of a chemical potential, which is given by
\begin{align}
    E_{N_\ua,N_\da} = E^0_{N_\ua,N_\da} - N_\ua\mu_\ua - N_\da\mu_\da.
\end{align}
If we can tune $\mu_\sigma$ to critical values $\mu_\sigma^c$ such that $E_{N_\ua + 1, N_\da + 1} = E_{N_\ua, N_\da}$, that is
\begin{align}
    E^0_{N_\ua+1,N_\da+1} - E^0_{N_\ua,N_\da} = \mu^c_\ua + \mu^c_\da,
    \label{meq}
\end{align}
then worm algorithms can sample worldline configurations in the particle-number sectors 
$(N_\ua,N_\da)$ and $(N_\ua+1,N_\da+1)$ very efficiently even at very large values of $\beta$. We can monitor this by computing the average particle number,
\begin{align}
    \langle N_\sigma \rangle = \frac{1}{Z_\mu}\ \Trace \Big( \sum_i \ N_{\sigma,i}\ e^{-\beta H_\mu}\Big),
    \label{eq:avgN}
\end{align}
and making sure that it fluctuates between the sectors $(N_\ua,N_\da)$ and $(N_\ua+1,N_\da+1)$ even when $\beta$ is large \cite{PhysRevD.81.125007}. Such fluctuations are crucial to the efficiency of our algorithm.

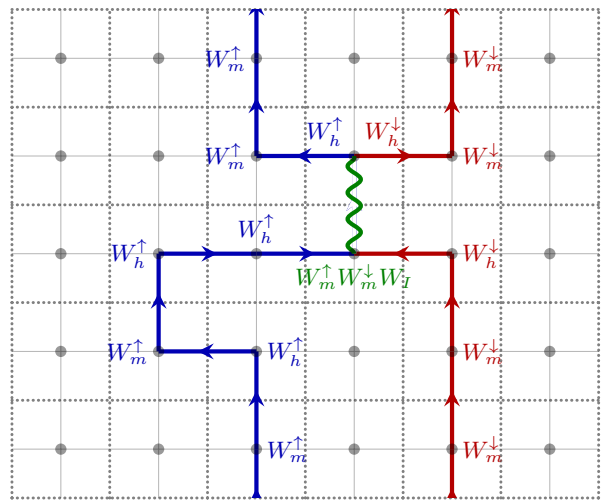
\begin{figure}[b]
\newcommand\FigWeightMove[1]{W^{#1}_m}
\newcommand\FigWeightHop[1]{W^{#1}_h}
\newcommand\FigWeightInteraction{W_I}

\tikzset{
snake it/.style={decorate, decoration=snake}}

\tikzset{
    dots/.style args={#1per #2}{%
      line cap=round,
      dash pattern=on 0 off #2/#1
    },
    inner grid/.style = {ultra thin, step=1.0, opacity=0.3},
    outer grid/.style = {line width=0.4mm, step=1.0, black, dots=15 per 1cm, opacity=0.5},
    site/.style = {circle, inner sep=0mm, minimum size=1.5mm, fill=black, opacity=0.4},
}

\def\InteractionColor{green!50!black}

% \scalebox{1}{
  \begin{tikzpicture}[
    % style to add an arrow in the middle of a path
    mid arrow/.style={postaction={decorate,decoration={
          markings,
          mark=at position .6 with {\arrow[#1]{stealth}}
        }}},
    weight label down/.style = {right,},
    % weight label down/.style = {, right,},
    weight label up/.style = {left,},
    node distance=0.5mm,
    label distance=0.5mm,
    % every node/.append style={
    %   font=\fontsize{6}{0}, },
    scale=1.3,
    worldline blue/.style = {ultra thick, color=blue!70!black, opacity=1.0},
    worldline red/.style = {ultra thick, color=red!70!black, opacity=1.0},
    worldline interaction/.style = {\InteractionColor, ultra thick },
    ]

    % Inner grid
    \begin{scope}
      \clip(-2,0) rectangle (4,5);
      \draw[xshift=0.5cm, yshift=0.5cm, inner grid] (-2.5,-0.5) grid (4,5);
    \end{scope}
    
    % Outer grid
    \draw[outer grid] (-2,0) grid (4,5);

    \begin{scope}
      \clip(-2,0) rectangle (4,5);

      %% Blue particle
      \draw [worldline blue]
      (0.5,-0.5) edge[mid arrow] (0.5,0.5) (0.5,0.5) node [weight label up, right] {$\FigWeightMove{\uparrow}$} 
      coordinate[] edge[mid arrow] ++(0,1) ++(0,1)  node [weight label up, right] {$\FigWeightHop{\uparrow}$} 
      coordinate[] edge[mid arrow] ++(-1,0)++(-1,0) node [weight label up, ] {$\FigWeightMove{\uparrow}$} 
      coordinate[] edge[mid arrow] ++(0,1) ++(0,1) node [weight label up] {$\FigWeightHop{\uparrow}$} 
      coordinate[] edge[mid arrow] ++(1,0) ++(1,0) node [weight label up, above] {$\FigWeightHop{\uparrow}$} 
      coordinate[] edge[mid arrow] ++(1,0) ++(1,0) node [weight label up, ] {}
      coordinate[] edge[mid arrow] ++(0,1)  ++(0,1)  node [weight label up, above left] {$\FigWeightHop{\uparrow}$} 
      coordinate[] edge[mid arrow] ++(-1,0) ++(-1,0) node [weight label up] {$\FigWeightMove{\uparrow}$} 
      coordinate[] edge[mid arrow] ++(0,1)  ++(0,1)  node [weight label up] {$\FigWeightMove{\uparrow}$} 
      coordinate[] edge[mid arrow] ++(0,1)  ++(0,1)  node [weight label up] {$\FigWeightMove{\uparrow}$} ;

      %% Red particle

      \draw [worldline red]
      (2.5, -0.5) edge[mid arrow] (2.5,0.5) (2.5, 0.5) node [weight label down] {$\FigWeightMove{\downarrow}$}
      coordinate [] edge[mid arrow] ++(0,1) ++(0,1) node [weight label down] {$\FigWeightMove{\downarrow}$}
      coordinate [] edge[mid arrow] ++(0,1) ++(0,1) node [weight label down] {$\FigWeightHop{\downarrow}$}
      coordinate [] edge[mid arrow]++(-1,0) ++(-1,0) node [weight label down, below, \InteractionColor] {$\FigWeightMove{\uparrow} \FigWeightMove{\downarrow} \FigWeightInteraction$} 
      coordinate [] edge[mid arrow, draw=white] ++(0,1) ++(0,1) node [weight label down, above right] {$\FigWeightHop{\downarrow}$}
      coordinate [] edge[mid arrow] ++(1,0) ++(1,0) node [weight label down] {$\FigWeightMove{\downarrow}$}
      coordinate [] edge[mid arrow] ++(0,1) ++(0,1) node [weight label down] {$\FigWeightMove{\downarrow}$}
      coordinate [] edge[mid arrow] ++(0,1) ++(0,1) node [weight label down] {$\FigWeightMove{\downarrow}$};

    %   \draw [worldline interaction] (1.5,2.5) edge[mid arrow, snake it] ++(0,1);
      \draw [worldline interaction] (1.5,2.5) edge[ snake it] ++(0,1);
    \end{scope}
    
    %% Draw nodes at each site
    \foreach \x in {-1.5,...,3.5} {
      \foreach \y in {0.5,...,4.5} {
        \node [site] (N-\x-\y) at (\x,\y) {};
      }
    }

  \end{tikzpicture}
\caption{An illustration of a worldline configuration ${\cal C}$ with $N_\ua=1$ and $N_\da=1$. The dots represent space-time lattice sites and the bold solid lines show the worldlines of the two particles. The interaction between the particles is shown as a wiggly temporal bond. The Boltzmann weights associated with lattice sites containing particles are also shown.}
\label{fig:wlconfig}
\end{figure}

\section{Worldline Formulation}
\label{sec:worldline-formulation}

Let us now construct the worldline formulation of the problem. We first write the Hamiltonian as $H_\mu = H_d - H_h$, a sum of a diagonal term and a hopping term, where 
\begin{align}
    H_d &= \sum_{i,\sigma} \Big(2 d \Hopping{\sigma} - \mu_\sigma\Big) N_{\sigma,i} + U  N_{\ua, i} N_{\da, i}, \\
    H_h &= \sum_{i,\hat\alpha,\sigma} \Hopping{\sigma} (c^\dagger_{i,\sigma} c_{i+\hat{\alpha},\sigma}^{\phantom{\dagger}} + c^\dagger_{i+\hat{\alpha},\sigma} c_{i,\sigma}^{\phantom{\dagger}}).
\end{align}
We then expand the partition function as
\begin{align}
\label{eq:zcontinuous}
  Z_\mu = &\sum_k\ \int_0^\beta\!\! dt_k \int_0^{t_k}\!\!  dt_{k-1} \cdots \int_0^{t_2}\!\!  dt_1 \\
        &\Trace\Bigg(
         e^{-(\beta-t_k) H_d} H_h
         e^{-(t_k-t_{k-1}) H_d}H_h \cdots H_h 
         e^{-t_1 H_d}\Bigg),
         \nonumber
\end{align}
which can be viewed as a hopping parameter expansion in continuous time. Since we do not truncate the sum over $k$ in \cref{eq:zcontinuous}, there is no approximation involved. Such an approach to write partition functions in continuous time is well known for hard-core bosons \cite{PhysRevLett.77.5130} and fermions \cite{PhysRevB.72.035122}. 
However,  to develop the worm algorithm,
it is convenient to discretize the time integrals by dividing $\beta$ into $L_T$ imaginary time steps of width $\varepsilon$, such that $\beta= \varepsilon L_T$. 
If we then compute the trace in the occupation number basis, we can approximate the partition function as a sum over worldline configurations ${\cal C}$ of both species of particles on a space-time lattice. 
We write this as 
\begin{align}
    Z_\mu = \sum_{\cal C} \ \Omega ({\cal C}),
    \label{zboson}
\end{align}
where $\Omega({\cal C})$ is the Boltzmann weight of each worldline configuration. \Cref{fig:wlconfig} gives an illustration of ${\cal C}$ on a $1+1$ dimensional space-time lattice. 
In the next section, we construct a worm algorithm to update such worldline configurations.
We usually perform calculations at several values of $\varepsilon$ and then extrapolate to the continuous time limit. 
We always find that the time-discretization errors are linear in $\varepsilon$ at leading order (see \namecrefs{fig:highdim} \labelcref{fig:highdim} and \labelcref{fig:epsilon-extrapolation-d3-U2}). 
In principle, this extra work can be avoided by directly taking the continuous time limit of the worm algorithm itself \cite{PhysRevLett.77.5130}.

The Boltzmann weights $\Omega({\cal C})$ can be obtained as a product of local weights associated to each space-time lattice site,
\begin{align}
    \Omega({\cal C}) = \prod_{i,t} W^{i,t}({\cal C}),
\end{align}
where $W^{i,t}({\cal C})$ is obtained from the local worldline configuration ${\cal C}$ at the space-time site $(i,t)$ and is a product of either $W^{\sigma}_e, W^{\sigma}_h$ or $W^{\sigma}_m$ for each species and  $\WeightInt$ that takes into account particle interactions. The allowed configurations at a site for each particle species are shown in \cref{fig2}. Either there is no particle or one particle that comes into the site and leaves it. If there is no particle, the site weight for that species is $W^{\sigma}_e=1$. On the other hand, when there is a particle, we choose the weight to depend only on its outgoing direction. If the particle hops to the neighboring spatial site, the weight is
\begin{align}
    W^{\sigma}_h = \Hopping{\sigma}\varepsilon.
    \label{wth}
\end{align}
If, instead, the particle moves forward in time, the weight is
\begin{align}
    W^{\sigma}_m = \exp\big(-\varepsilon (2 d \Hopping{\sigma}- \mu_\sigma)\big).
    \label{wtm}
\end{align}
The interaction among the particle species is taken into account through the weight
\begin{align}
\WeightInt = \exp\left(-\varepsilon U\right),
\label{wti}
\end{align}
if both particles hop forward in time together. 
Otherwise we set $\WeightInt=1$.  
With these definitions, we can express the weight of a configuration as
\begin{align}
    \Omega(\MCConfig) &= 
    (\WeightMove^\uparrow)^{\LT\NumParticles{\uparrow}}
    (\WeightMove^\downarrow)^{\LT\NumParticles{\downarrow}}
    (\WeightHop^\downarrow)^\NumHops{\downarrow}
    (\WeightHop^\uparrow)^\NumHops{\uparrow}
    (\WeightInt)^\NumInteractions,
    \label{eq:weight-config}
\end{align}
where $\NumParticles{\sigma}$ is the number of $\sigma$-particles in the configuration $\MCConfig$, $\NumHops{\sigma}$ is the number of hops and $\NumInteractions$ is the number of interacting temporal bonds. 
\begin{figure}[hbt]
\centering
\WormBox{xxx}~\WormBox{xlu}~\WormBox{xru}\\
\WormBox{xdl}~\WormBox{xdr}~\WormBox{xdu}\\
\WormBox{xlr}~\WormBox{xrl} \\
\caption[Begin/End updates]{The allowed worldline configurations ${\cal C}$ on a 1+1 dimensional space-time lattice site for one particle species.
Using the weights defined in 
\crefrange{wth}{wti},
the weights of the configurations from left to right are (top row) $1,\WeightMove,\WeightMove$, (middle row) $\WeightHop,\WeightHop,\WeightMove$, (bottom row) $\WeightHop,\WeightHop$. In addition to these weights if both layers have the weight $\WeightMove$ we multiply the product with $\WeightInt$.}
\label{fig2}
\end{figure}

\newcommand\WormGroupSpace{\hskip0.2in}
\newcommand\WormSpace{\hskip-0.2in}

\begin{figure*}[htbp]
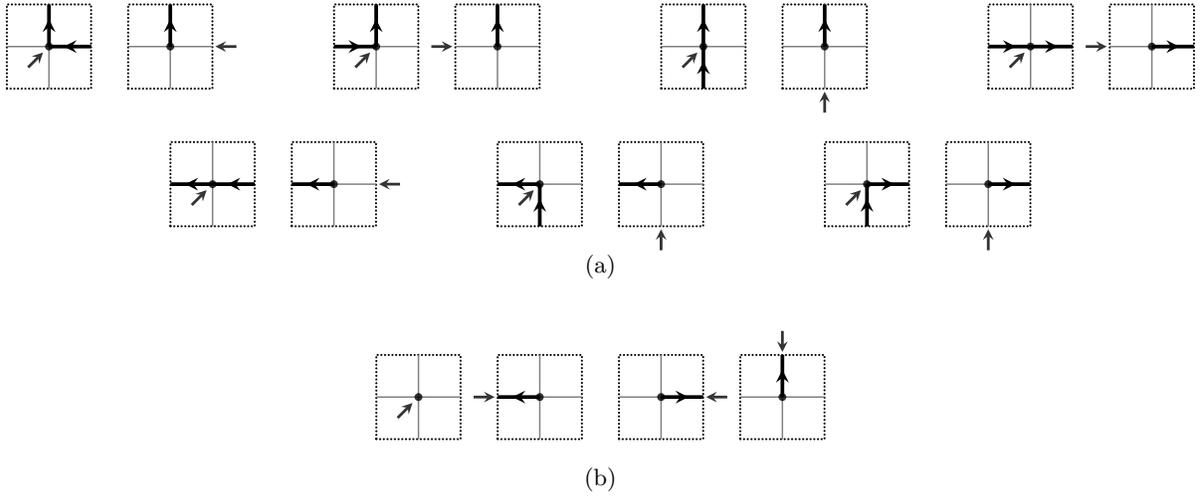

\WormBox{cru}\WormSpace\WormBox{rxu} \WormGroupSpace
\WormBox{clu}\WormSpace\WormBox{lxu} \WormGroupSpace
\WormBox{cdu}\WormSpace\WormBox{dxu} \WormGroupSpace 
\WormBox{clr}\WormSpace\WormBox{lxr} \WormGroupSpace \\
\WormBox{crl}\WormSpace\WormBox{rxl} \WormGroupSpace
\WormBox{cdl}\WormSpace\WormBox{dxl} \WormGroupSpace 
\WormBox{cdr}\WormSpace\WormBox{dxr} \WormGroupSpace \\
(a)\\[2em]
\WormBox{cxx}\WormSpace\WormBox{lxl}\WormSpace\WormBox{rxr}\WormSpace\WormBox{uxu}\\
(b)
\caption{The eight groups of possible begin-end updates in $1+1$ dimensions, classified by where the worm update begins: (a) on a site with an existing particle, or (b) on an empty site. The small arrow in each local configuration depicts the direction of an incoming worm.  The outgoing worm direction can be obtained by reversing this arrow. \label{fig:worm-update-begin-end}}
\end{figure*}

It is possible to compute a variety of observables easily in the worldline formulation. 
For example, the average energy defined in \cref{eq:avgE} can be expressed in the worldline formulation, up to $O(\varepsilon)$ errors, as
\begin{align}
    \langle E \rangle &=  \frac{1}{Z_\mu} \sum_{\MCConfig} E(\MCConfig) \ \Omega(\MCConfig)
    \label{eq:wlenergy}
\end{align}
where 
$E(\MCConfig)$ is the energy of a worldline configuration ${\MCConfig}$, defined as
\begin{align}
E(\MCConfig) &= U \frac{\NumInteractions(\MCConfig) }{L_T} +  \sum_{\sigma} \left[ - \frac{\NumHops{\sigma}(\MCConfig) }{{\beta}} + 2d \Hopping{\sigma} \NumParticles{\sigma}({\cal C}) \right].
\label{eq:energy-of-worldline}
\end{align}
This expression for $E(\MCConfig)$ can be derived from \cref{eq:weight-config} by taking the appropriate derivative with respect to $\beta$.
The average particle number for each species defined in \cref{eq:avgN} is straightforward since worldline configurations have a well-defined particle number $N_\sigma(\MCConfig)$. Thus, we get 
\begin{align}
\langle N_\sigma\rangle = \frac{1}{Z_\mu} \sum_{\cal C} N_\sigma(\MCConfig) \ \Omega(\MCConfig).
\end{align}
We can also measure ratios of partition functions, like the one we define later in \cref{zratio}, by designing an appropriate sampling method during the Monte Carlo update.

\begin{figure*}[hbt]
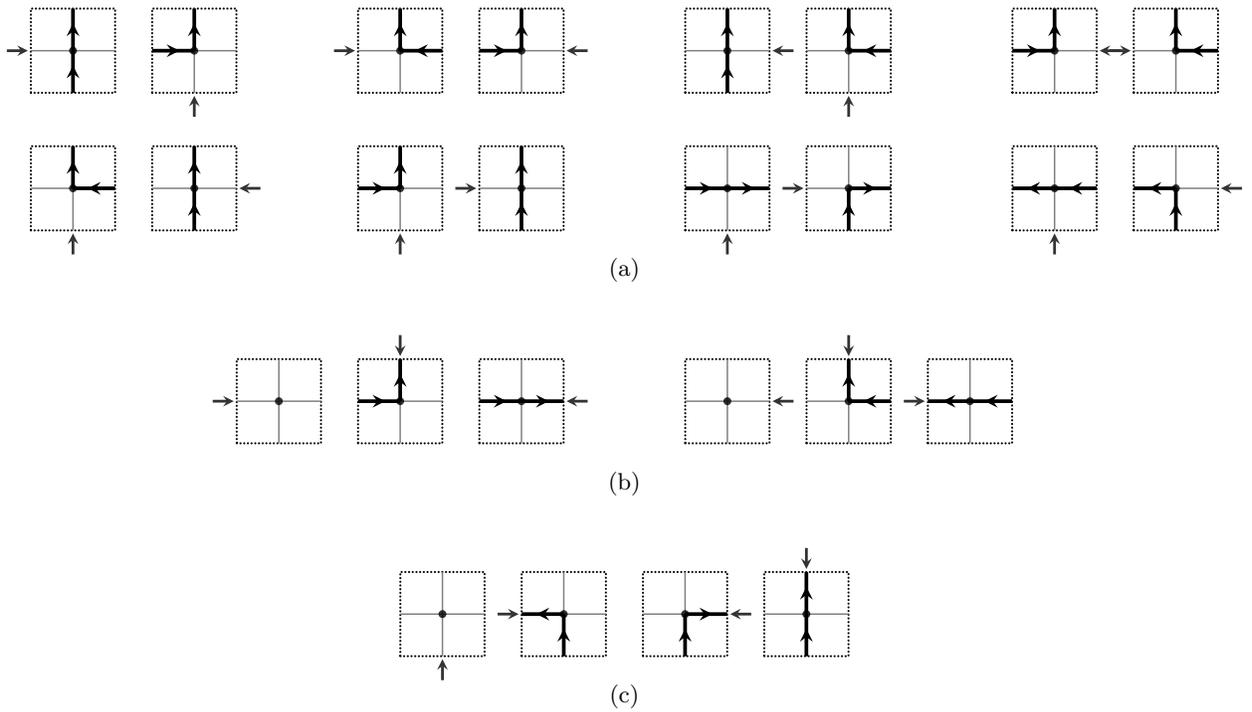

\WormBox{ldu}\WormSpace\WormBox{dlu} \WormGroupSpace
\WormBox{lru}\WormSpace\WormBox{rlu} \WormGroupSpace
\WormBox{rdu}\WormSpace\WormBox{dru} \WormGroupSpace
\WormBox{rlu}\WormSpace\WormBox{lru} \WormGroupSpace \\
\WormBox{dru}\WormSpace\WormBox{rdu} \WormGroupSpace
\WormBox{dlu}\WormSpace\WormBox{ldu} \WormGroupSpace
\WormBox{dlr}\WormSpace\WormBox{ldr} \WormGroupSpace
\WormBox{drl}\WormSpace\WormBox{rdl}\\
(a) \\[2em]
\WormBox{lxx}\WormSpace\WormBox{ulu}\WormSpace\WormBox{rlr}
\WormGroupSpace
\WormBox{rxx}\WormSpace\WormBox{uru}\WormSpace\WormBox{lrl}\\
(b) \\[2em]
\WormBox{dxx}\WormSpace\WormBox{ldl}\WormSpace\WormBox{rdr}\WormSpace\WormBox{udu} \\
(c) \\
\caption{Local worm updates in $1+1$ dimensions when (a) the worm enters a site which already contains a particle worldline;  (b) the worm enters an empty site along the spatial direction; or (c)  the worm enters an empty site along the forward time direction. \label{fig:worm-update-midway}}
\end{figure*}

\section{The Worm Algorithm}
\label{sec:worm-algorithm}

It is possible to develop worm algorithms to update the worldline configurations ${\MCConfig}$ of the type shown in \cref{fig:wlconfig} \cite{Adams:2003cca,chandrasekharan:2006tz}. 
During the worm update, we pick each particle species and update its worldline configuration while keeping the worldline of the other species fixed. 
To perform the update, we pick a space-time site at random and create a defect in the worldline configuration in the form of either a particle creation or annihilation event. We choose the worm head as the position of the annihilation operator and the tail as the position of the creation operator. We then move the head on the space-time lattice, while keeping the location of the tail fixed, using local moves that obey detailed balance. The worldline configuration gets updated on each site the worm visits. The update ends when the worm head meets the tail 
and removes the defect. The particle number can be monitored during the worm update and local rules can be chosen so as to sample particle numbers within a fixed range.

For a given worldline configuration, we define the local configuration at a space-time site as the incoming and the outgoing directions of the particle. In addition we also include the information about how the worm enters or leaves that site. Each box in \cref{fig:worm-update-begin-end,fig:worm-update-midway} represents such a local configuration with the worm entering the site. Local configurations with the worm leaving the site can be constructed from these by reversing the direction of the worm arrow. 
We then group configurations that can transform into each other under local rules which satisfy detailed balance.

To understand this procedure better, let us consider local updates that begin or end a worm update, shown in \cref{fig:worm-update-begin-end} for a $1+1$ dimensional lattice.
When the worm update begins, the incoming worm direction is shown as a diagonal arrow entering the site. 
The outgoing worm direction could be along any of the neighboring space-time lattice sites as long as that move is allowed. 
For detailed balance to work, the worm update should also be allowed to end through the reverse process. There are two classes of such begin-end updates based on whether the first site contains a particle or not.  If the site contains a particle (\cref{fig:worm-update-begin-end}a), then a creation operator is introduced on the site (i.e., the site becomes the worm tail) and the worm head that contains the annihilation operator moves to the neighboring site through which the particle came to the first sit (i.e., with probability one the outgoing direction is chosen to be the backward direction the worldline). For detailed balance to be satisfied the reverse process is also chosen with probability one (i.e., when the worm head enters the site that contains the tail, the worm update ends). Thus, these two local pair of configurations are grouped together. There are seven possible such pairs of begin-end updates in $1+1$ dimensions, as shown in \cref{fig:worm-update-begin-end}a. In higher dimensions there would be more.

It is also possible that the worm update begins at an empty site (\cref{fig:worm-update-begin-end}b). 
In such a case, a new particle is proposed to be created on that site (i.e, a creation operator is placed on the site which becomes the worm tail) and the worm head moves either to one of the neighboring spatial sites or upwards to the neighboring temporal site. 
In $1+1$ dimensions, the empty site can thus transform into three possible local configurations. These four configurations are grouped together and shown in \cref{fig:worm-update-begin-end}b. 
We assign probabilities for moves within this set of four local configurations such that detailed balance is satisfied.  Details on how these probabilities are chosen are discussed below.

Between the begin and end updates, the worm head moves through space-time lattice sites.
Such moves can be classified into three classes, as shown in \cref{fig:worm-update-midway}. The simplest class involves a move where the worm enters the site that already has a particle on it. In this case, the entering direction of the worm cannot be the same as the incoming particle direction.
Thus, with probability one, the outgoing worm direction can be exchanged with the incoming particle direction.
Eight such pairs of configurations exist in $1+1$ dimensions and are shown in \cref{fig:worm-update-midway}a. 
The next class involves a worm moving into an empty site from a spatial neighbor. 
Then the worm can exit through the forward time direction or through a different spatial direction. This groups $2d+1$ local configurations together in $d$ spatial dimensions. 
The two possible groups of three configurations in $1+1$ dimensions are shown in \cref{fig:worm-update-midway}b. 
The third class involves a worm entering into an empty site along the forward time direction. Then the outward worm direction could be along any of the spatial or forward time directions. 
In $d$ dimensions, there are $2d+2$ such local configurations that can transform into each other.
The only possible such group of four configurations in $1+1$ dimensions is shown in \cref{fig:worm-update-midway}c.

For the efficiency of the worm method, probabilities for moving the worm head must be chosen so as to avoid a ``bounce'' as much as possible \cite{PhysRevE.66.046701}. A bounce occurs when the local configuration does not change (i.e, the probability to simply reverse the incoming worm direction wins). Note that bounces can be completely eliminated among the pairs of configurations in \namecrefs{fig:worm-update-begin-end}~\ref{fig:worm-update-begin-end} and \ref{fig:worm-update-midway}, since they have the same weight. 
In these cases, the local worm update simply toggles the two. 
However, in the case of groups that contain more than two local configurations, like the four-configuration group in \cref{fig:worm-update-begin-end}b or the three- and four-configuration groups in Figs.~\ref{fig:worm-update-midway}b and \ref{fig:worm-update-midway}c, we need to make sure the worm bounces are minimized as much as possible. For small values of $\varepsilon$, since two of the weights are very close to one, and spatial hops are rare this is almost always possible. 

Let $P_{ab}$ be the transition probability to go from local configurations $a$ to $b$.
To illustrate how we choose $P_{ab}$ to satisfy detailed balance, we consider the example of a three-configuration group in \cref{fig:worm-update-midway}b. The bounce probability is given by
\begin{align}
P_{aa} = 1 - \sum_{b \neq a} P_{ab}.
\end{align}
The weights of the three configurations are $W_1=1$, $W_2=\WeightMove$ (or $W_2 = \WeightMove\WeightInt$) and $W_3=\WeightHop$. 
In this case, $W_1$ and $W_2$ are close to one, while $W_3$ is of the order of $\varepsilon$. 
There are three possible orderings of the weights: Case~I: $W_2 > W_1 + W_3$, Case~II: $W_1 > W_2 + W_3$ and Case~III: $W_a \leq W_b + W_c$ for all $a,b,c$ different. In \cref{tab:transition-prob-1}, we give the transition probabilities $P_{ab}$ for all these cases.

\begin{table}
  \SetTableProperties{}
  \begin{tabular}{c @{\hspace{0.8em}}|@{\hspace{0.8em}} c}
    \TopRule
    \multicolumn{2}{c}{Case I: $W_2 > W_1 + W_3$} \\
    \multicolumn{2}{c}{$P_{21} = W_1/W_2$, $P_{23} = W_3/W_2$,$P_{12} = P_{32} = 1$} \\[5pt]
    \MidRule
    \multicolumn{2}{c}{Case II: $W_1 > W_2 + W_3$} \\
    \multicolumn{2}{c}{
    $P_{12} = W_2/W_1$, $P_{13} = W_3/W_1$,
    $P_{21} = P_{31} = 1$} \\[5pt]
    \MidRule
    \multicolumn{2}{c}{Case III: $W_a \leq W_b + W_c$}\\
    $P_{12} = (W_1+W_2-W_3)/2W_1$ & $P_{13} = (W_1+W_3-W_2)/2W_1$ \\
    $P_{21} = (W_2+W_1-W_3)/2W_2$ & $P_{23} = (W_2+W_3-W_1)/2W_2$ \\
    $P_{31} = (W_3+W_1-W_2)/2W_3$ & $P_{32} = (W_3+W_2-W_1)/2W_3$ \\
    \BotRule
  \end{tabular}
  \caption{Table of transition probabilities among the group of three configurations in \cref{fig:worm-update-midway}b \label{tab:transition-prob-1}}
\end{table}

\begin{table}\SetTableProperties{}
  \begin{tabular}{c @{\hspace{0.2em}}|@{\hspace{0.2em}} c}
    \TopRule
    \multicolumn{2}{c}{Case I: $W_2 > W_1 + W_3 + W_4$} \\
    \multicolumn{2}{c}{$P_{21} = W_1/W_2$, $P_{23} = W_3/W_2$,$P_{24} = W_4/W_2$} \\
    \multicolumn{2}{c}{$P_{12} = P_{32} = P_{42} = 1$} \\
    \MidRule
    \multicolumn{2}{c}{Case II: $W_1 > W_2 + W_3 + W_4$} \\
    \multicolumn{2}{c}{$P_{12} = W_2/W_1$, $P_{13} = W_3/W_1$,$P_{14} = W_4/W_1$} \\
    \multicolumn{2}{c}{$P_{21} = P_{31} = P_{41} = 1$} \\
    \MidRule
    \multicolumn{2}{c}{Case III: $W_a \leq W_b + W_c$}\\
    $P_{12} = (W_1+W_2-2\WeightHop)/2W_1$ &
                                     $P_{13} = (W_1+2\WeightHop-W_2)/4W_1$ \\
    $P_{14} = (W_1+2\WeightHop-W_2)/4W_1$ &
                                     $P_{21} = (W_2+W_1-2\WeightHop)/2W_2$ \\
    $P_{23} = (W_2+2\WeightHop-W_1)/4W_2$ &
                                     $P_{24} = (W_2+2\WeightHop-W_1)/4W_3$ \\
    $P_{31} = (2\WeightHop+W_1-W_2)/4\WeightHop$ &
                                     $P_{32} = (2\WeightHop+W_2-W_1)/4\WeightHop$ \\
    $P_{41} = (2\WeightHop+W_1-W_2)/4\WeightHop$ &
                                     $P_{42} = (2\WeightHop+W_2-W_1)/4\WeightHop$ \\
    \multicolumn{2}{c}{$P_{34}=P_{43} = 1/2$} \\
    \BotRule
  \end{tabular}
  \caption{Table of transition probabilities among the group of four configurations in \namecrefs{fig:worm-update-begin-end}~\ref{fig:worm-update-begin-end}b and \ref{fig:worm-update-midway}c \label{tab:transition-prob-2}}
\end{table}

The above method of constructing transition probabilities is easily extended to other groups of local configurations in our work. For example the group of four configurations in \namecrefs{fig:worm-update-begin-end}~\ref{fig:worm-update-begin-end}b and \ref{fig:worm-update-midway}c have weights $W_1=1$, $W_2 = \WeightMove$ (or $W_2 = \WeightMove\WeightInt$) and $W_3 = W_4 = \WeightHop$. In these cases we can imagine the configurations with weights $W_3$ and $W_4$ as a single configuration with weight $W_3+W_4=2\WeightHop$ and again use the probabilities of \cref{tab:transition-prob-1}. The complete set of transition probabilities for this case is given in \cref{tab:transition-prob-2}. We test our algorithm for small lattices in $d=1,\,2,\,3$ dimensions and the results of these tests are given in \cref{sec:exact-results-small-lattice}.

\section{Fermions and Sign Problems}
\label{sec:sign}

Any lattice model of hard-core bosons can be converted into a model of fermions if we take into account the fermion permutation sign. The partition function for fermions $Z^{\text{f}}_\mu$, can be obtained from the bosonic one in \cref{zboson} through the relation 
\begin{align}
    Z^{\text{f}}_\mu = \sum_{\cal C} {S}({\cal C}) \Omega({\cal C})
    \label{zfermion}
\end{align}
where $S(\cal C)$ is the fermion permutation sign of the worldline configuration ${\cal C}$.  If we define the average fermion permutation sign $\langle S\rangle$ as the ratio of the two partition functions in a fixed particle number sector as
\begin{align}
    \langle S \rangle \ = \ \frac{Z^{\text{f}}_\mu}{Z_\mu},
\end{align}
then the energy observable in the fermionic theory can be defined using the relation
\begin{align}
    \langle \EnergyFermionic  \rangle = \langle E \rangle - \frac{1}{\beta} \ln\left(\langle S \rangle\right),
    \label{eq:fermigse}
\end{align}
where $\langle E \rangle$ is the average energy for hard-core bosons computed through \cref{eq:avgE}. The ground-state energy of the theory with fermions is then obtained in the low temperature limit
\begin{align}
    E^{\text{f}\,0}_{N_\ua,N_\da} &= \lim_{\beta \rightarrow 0} \langle \EnergyFermionic  \rangle.
\end{align}
The situation simplifies in one spatial dimension.  It can be shown that with periodic boundary conditions and an odd number of particles of each species, $\langle S\rangle = 1$.  In the case of the open boundary conditions or in a trapping potential, the fermions cannot wind around the box and $\< S \> = 1$ again.
This implies that the fermionic and bosonic energies are the same, that is, $E^{0f}_{N_\ua,N_\da} = E^{0}_{N_\ua,N_\da}$. We use this result in the next two sections when we study one-dimensional problems.

\begin{figure}[htbp]
\includegraphics[width=\linewidth]{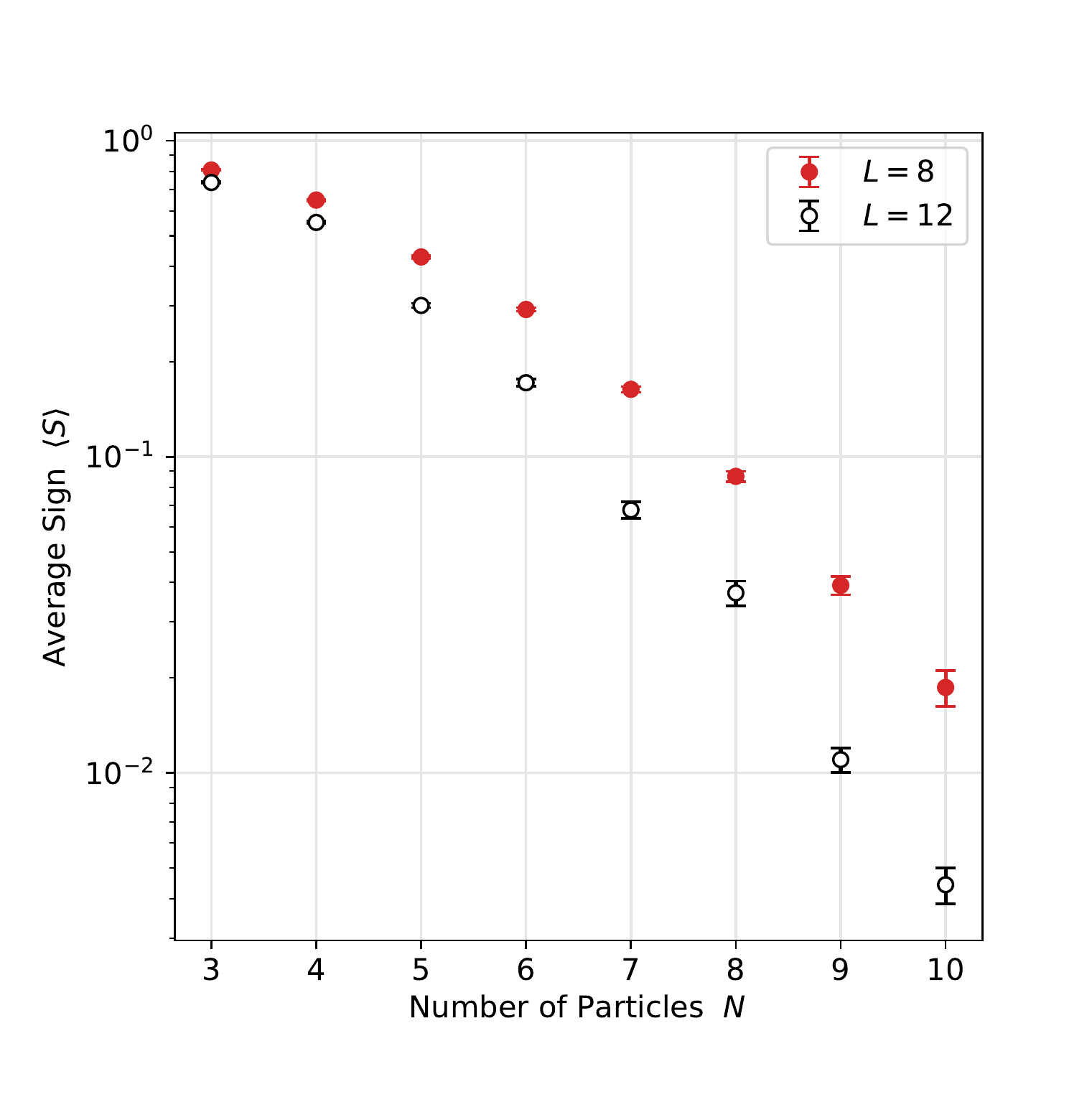}
\caption{The average sign $\< S \>$ in three spatial dimensions at two different lattice sizes as a function of $N=N_\ua+ N_\da$ in the symmetric repulsive model with $m_\sigma=1$ and $U=4.0$ at $\beta=5$ ($L=8$) and $12$ ($L=12$). The values of $\beta$ is chosen such that roughly $(2\pi/L)^2\beta \sim 1.5$.  The results are obtained a finite temporal lattice spacing of  $\varepsilon=0.01$.}
  \label{fig:sign}
\end{figure}

\begin{figure*}[htbp]
    \centering
    \includegraphics[width=\FigureWidth,valign=t]{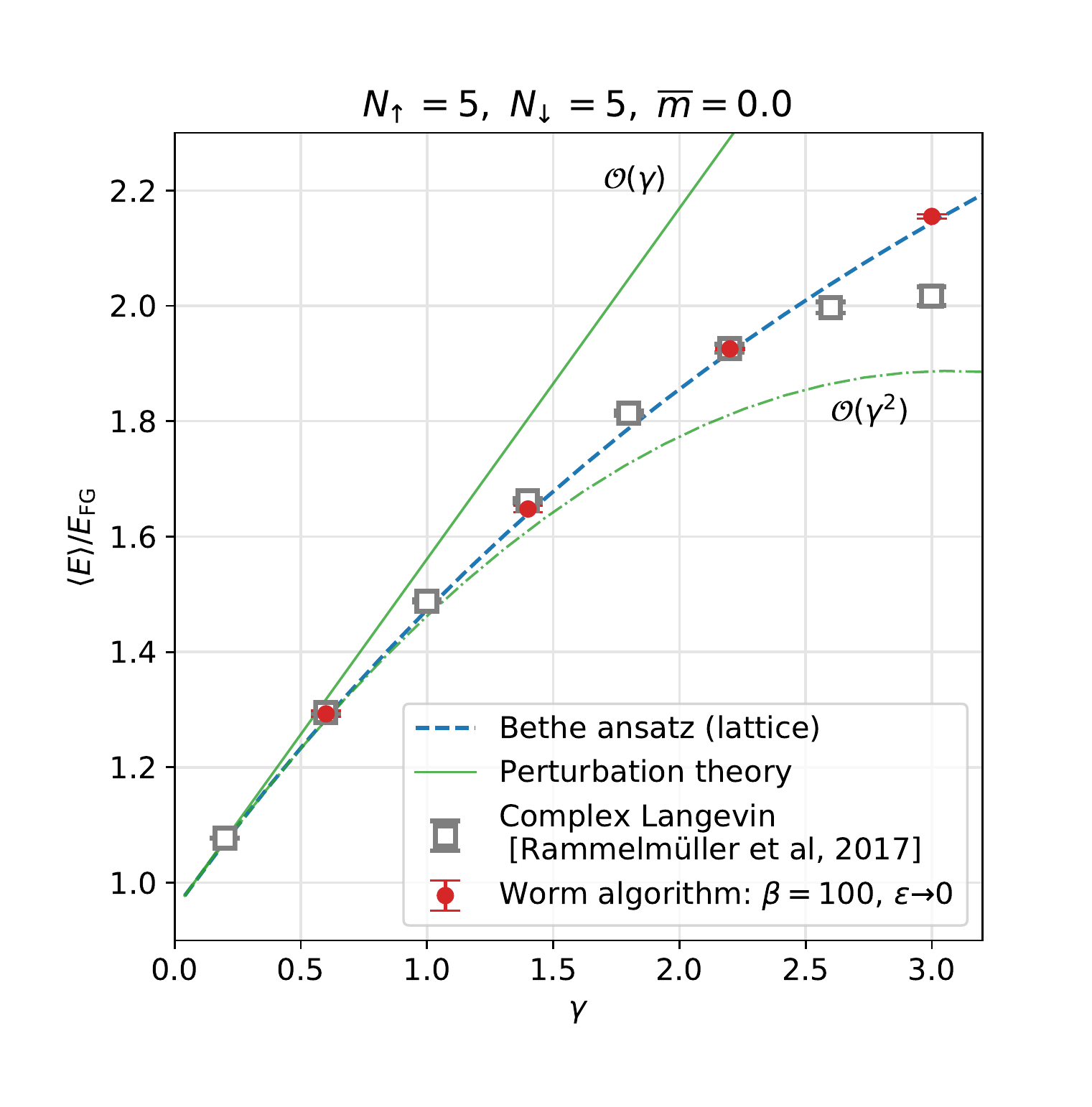}\hfill
    \includegraphics[width=\FigureWidth,valign=t]{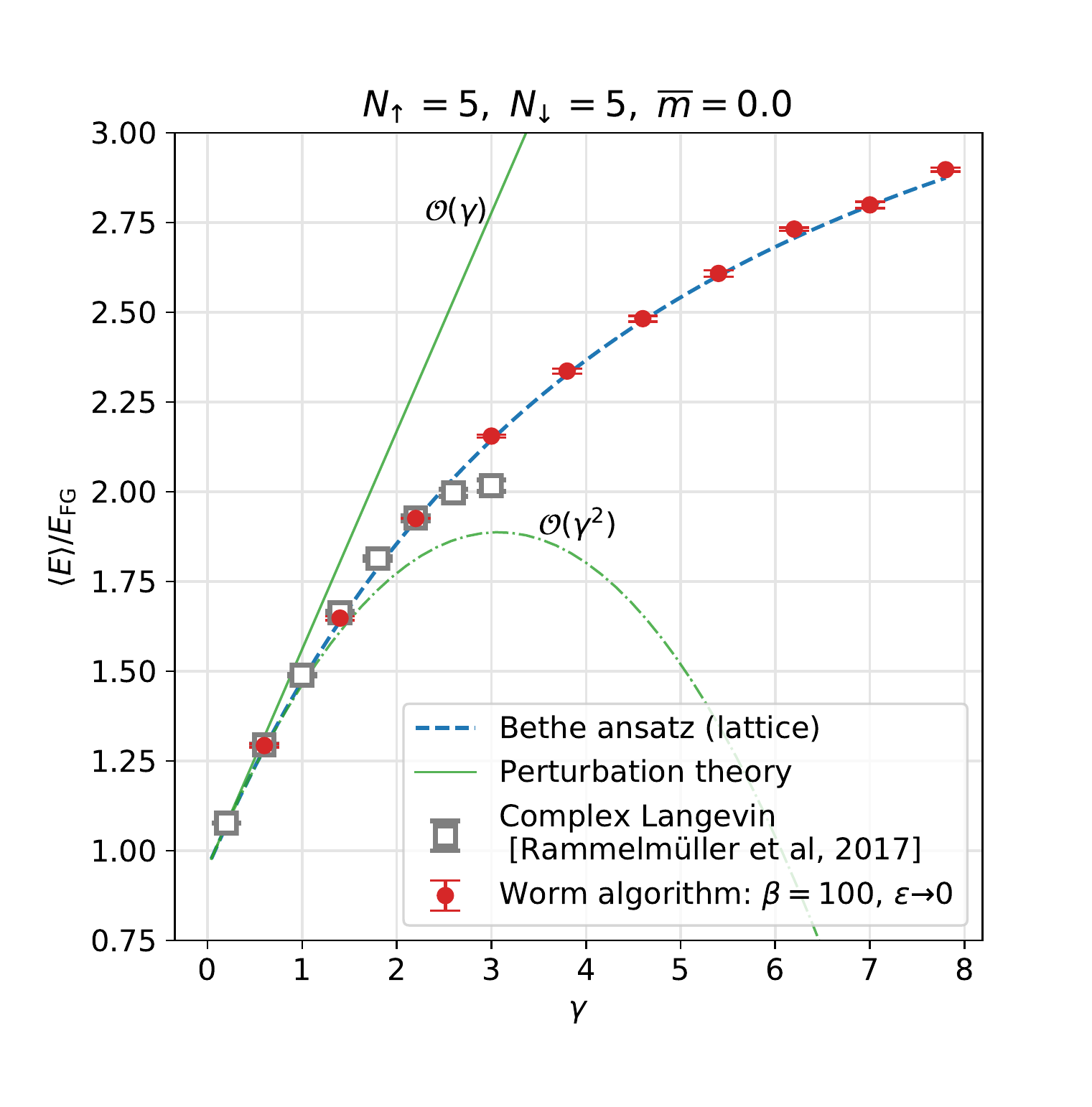}
    \caption{Plots of ground state energy as a function of $\gamma$ in the mass balanced case ($\mbar=0.0$), with $N_\ua=N_\da=5$. The left figure focuses on the perturbative regime while the right figure extends it to the non-perturbative region. The complex Langevin results shown are from \cite{PhysRevD.96.094506}. Note that they begin to disagree with exact results around $\gamma \gtrsim 2.5$.}
    \label{fig:bethe-ansatz}
\end{figure*}

In higher dimensions, we expect $\langle S\rangle < 1$.  In particular, the average permutation sign will suffer from a severe signal-to-noise ratio problem when $\beta$ becomes large and in the presence of large number of particles.
This is essentially the sign problem coming back to haunt the Monte Carlo method. However, it is interesting to note that our algorithm does does not distinguish the free problem ($U=0$) from the interacting cases of attraction ($U < 0$) or repulsion ($U > 0$). Our approach also does not differentiate between whether the two species have similar or different masses. The sign problem is severe in all cases, although we find it somewhat milder with attractive as compared to repulsive interactions. In \cref{fig:sign}, we plot the average sign $\langle S\rangle$ as a function of the particle number $N=N_\ua+N_\da$ in the mass-balanced repulsive model with $m_\sigma=1$ and $U=4$ at temperatures such that $\beta(1-\cos(2\pi/L)) \sim 1.5$. 
For even $N$ we choose $N_\ua = N_\da$ and for odd $N$ we use $N_\ua = N_\da+1$. 
We see that up to $N=10$ the sign problem is not very severe at these intermediate temperatures. 
This gives us confidence that, in combination with ideas of fermion bags \cite{PhysRevD.96.114502} and exponential error-reduction techniques \cite{Luscher:2001up}, we may be able to use this approach to study few-body physics even in higher dimensions. 
We show some preliminary evidence for this in \cref{sec:higher-dimensions}.

\begin{figure*}[htbp]
  \centering
  \includegraphics[width=\FigureWidth,valign=t]{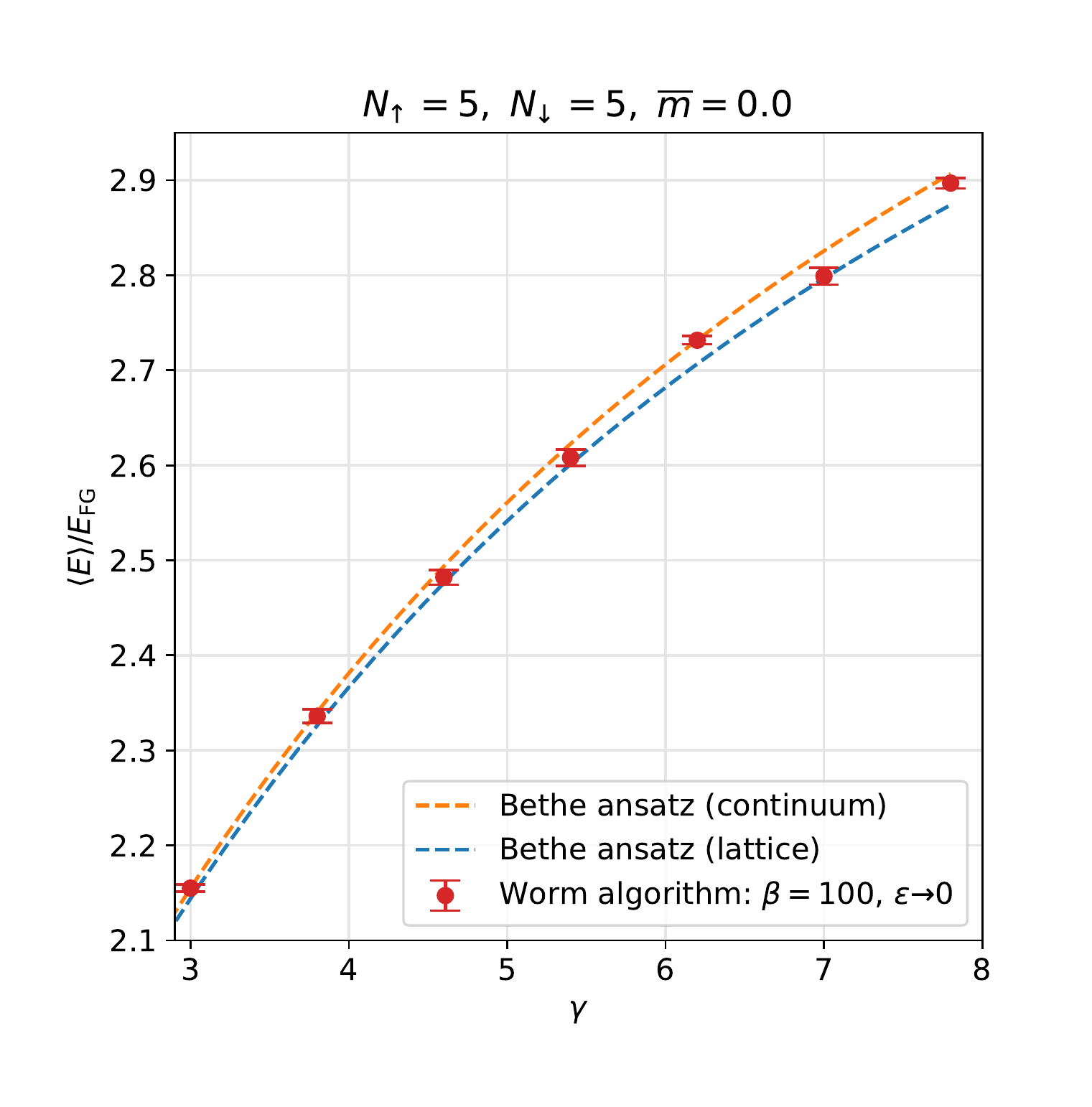}
  \hfill
  \includegraphics[width=\FigureWidth,valign=t]{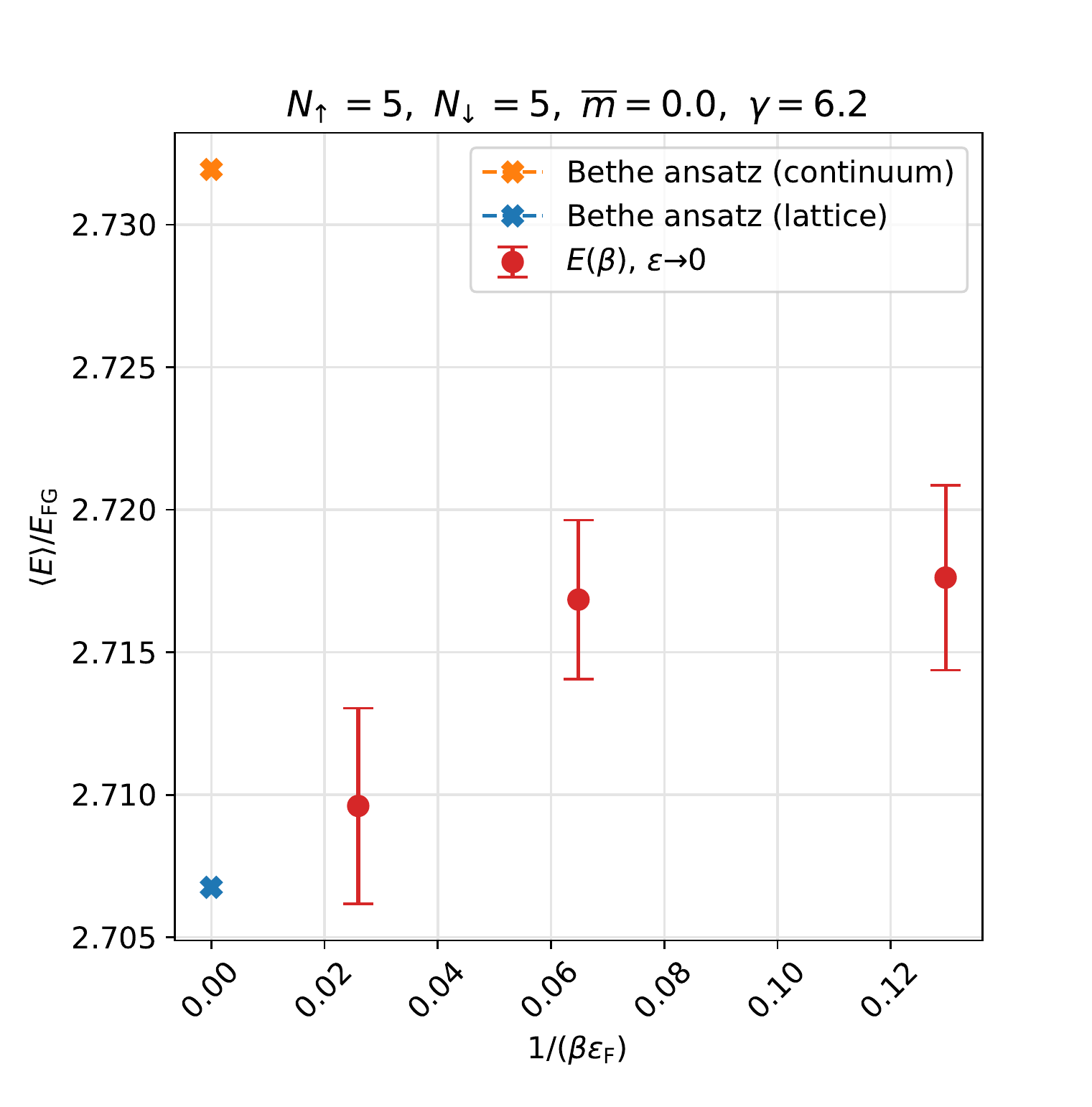}
  \caption{The left figure shows a comparison between the worm algorithm results at $\LX=40$, $\beta=100$, $\varepsilon \rightarrow 0$, $\mbar=0.0$,  $N_\ua=N_\da=5$, with the results obtained using the Bethe ansatz on a lattice with $\LX=40$ and in the continuum. The $\beta$ dependence of the energies plotted at $\gamma=6.2$ in the right figure shows that the worm algorithm results are sensitive to the small difference between the lattice and continuum. In the $\beta \EnergyFermi{} \to \infty$ limit, we recover the exact lattice result as expected.}
  \label{fig:bethe-ansatz-beta-extrapolation}
\end{figure*}

\begin{figure*}
\includegraphics[width=\FigureWidth,valign=t]{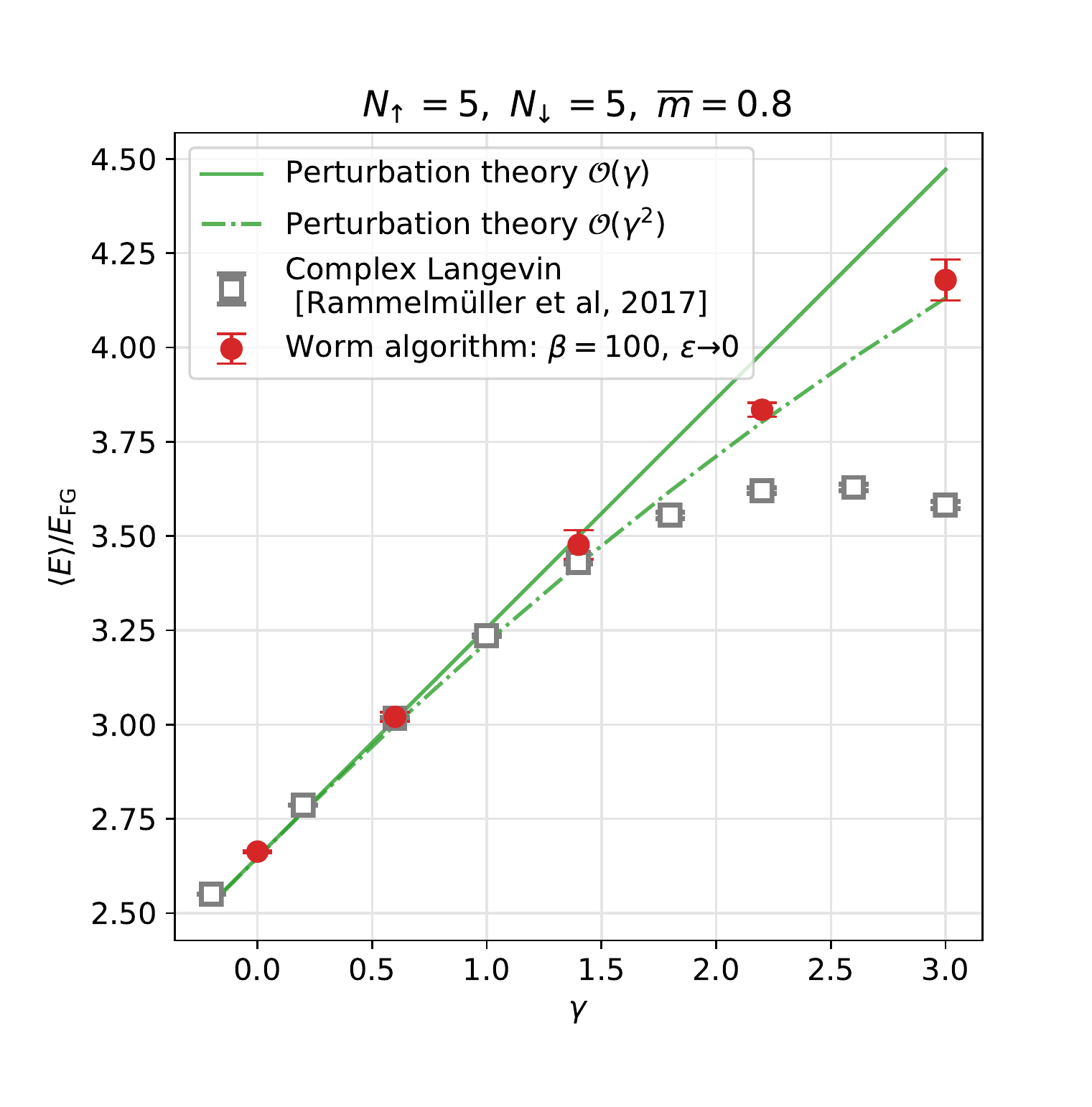}
\hfill
\includegraphics[width=\FigureWidth,valign=t]{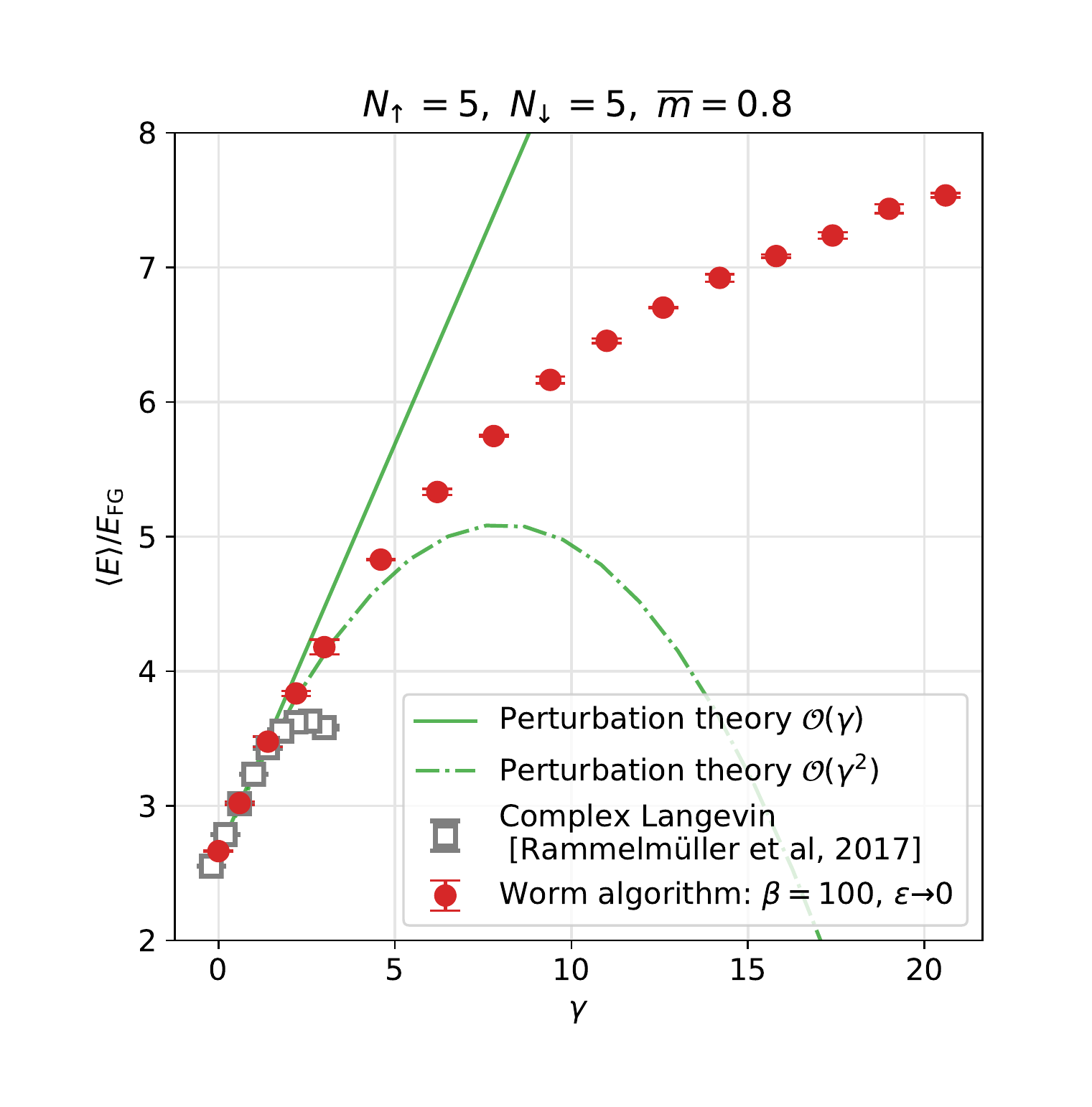}
\caption{Plots of ground state energy as a function of $\gamma$ in the high mass imbalanced case ($\mbar=0.8$), with $N_\ua=N_\da=5$. The left figure focuses on the perturbative regime while the right figure extends it to the non-perturbative region. The complex Langevin results shown are from \cite{PhysRevD.96.094506} and begin to disagree with the worm algorithm around $\gamma \gtrsim 2$, even in the regime where second-order perturbation theory works well.}
\label{fig:55-repulsive-mbar=0.8}
\end{figure*}

\section{One-dimensional Systems}
\label{sec:1d-results}

In this section, we report our results for a wide range of coupling strengths and mass-imbalances in one spatial dimension for spin-balanced systems ($N_\uparrow = N_\da$).  We fix the box size to be $\LX=40$ and impose periodic boundary conditions.  
As mentioned in the previous section, an odd number of fermions of each species is equivalent to hard-core bosons and our so method is directly applicable to fermions. 
We focus on computing the average energy $\langle E\rangle$ (see \cref{eq:avgE}) at a fixed $\beta=100$. In the next section, we discuss how our results change with $\beta$ for a few cases in more detail and how understanding this dependence is important to extract the ground-state energy accurately. We do perform an extrapolation to zero temporal lattice spacing $\varepsilon \to 0$, since these errors can be large. Details of this extrapolation are discussed further in  \cref{sec:higher-dimensions} and \cref{sec:exact-results-small-lattice}. 

We denote the total number of particles as $N = N_\ua + N_\da$ and the particle-number density by $n=N/\LX$. We define the average mass and the mass-imbalance parameter as
\begin{align}
  m = \frac{m_\uparrow + m_\downarrow}{2}, \quad\mbar = \frac{m_\uparrow - m_\downarrow}{m_\uparrow + m_\downarrow},
\end{align}
respectively. We work in units with $m_\sigma = 1$ and $a=1$ and parametrize the interaction strength by $\gamma$, which is related to the bare coupling $U$ in the lattice model \eqref{eq:lattice-model} by
\begin{align}
  \gamma = \frac{U}{n}.
\end{align}
To facilitate comparison with literature, we report all energies in units of the corresponding one-dimensional ideal spin- and mass-balanced Fermi-gas ground state energy $\EnergyFG$ in the continuum defined as
\begin{align}
  \EnergyFG = \frac13 N \EnergyFermi,
\end{align}
where $\EnergyFermi = \pi^2 n^2/8m$ is the Fermi energy.

\subsection{Mass-Balanced Systems}

The continuum model \eqref{cmodel} for fermions with $\mbar=0$ is known as the Gaudin-Yang model and can be solved exactly when $g > 0$ using nested Bethe ansatz \cite{yang_exact_1967, gaudin_systeme_1967}. 
This solution can also be extended to the lattice model \eqref{eq:lattice-model} in one spatial dimension ($d=1$) for an arbitrary lattice size $\LX$, when coupling $U \geq 0$, and number of particles $N = N_\uparrow + N_\downarrow$ as long as $m_\ua = m_\da$ \cite{lieb_absence_1968}. 
Since attractive models on a finite lattice can be related to repulsive models by a particle-hole transformation on one of the spins, we can compute the exact ground-state energies of our lattice model in one spatial dimension when $\mbar = 0$ for a wide range of the coupling strengths, both attractive and repulsive. 
This allows us to test our Monte Carlo method for the mass-balanced case against the exact solution. As we show below, we find excellent agreement between the two. 

For completeness, let us quickly review the main steps that go into the exact computation of the ground-state energy.  
We discuss the solution for the lattice model only, since the solution in the continuum can easily be obtained from it.  
Let us assume we have $N_\downarrow=M$ spin-down fermions, and $N_\uparrow = N-M$  spin-up fermions.  Let us normalize our energies with $t_\uparrow = t_\downarrow = t$.  Let $p_i = (p_1, \dotsc, p_N)$ and $\lambda_\alpha = (\lambda_1, \dotsc, \lambda_M)$ be two sets of ascending real numbers. Then the following $N+M$ coupled non-linear equations in the $N+M$ variables $\{p_i, \lambda_\alpha\}$ determine the complete spectrum of this system:
\begin{align}
  2\pi I_i &= \LX p_i - \sum_{\alpha=1}^{M} \theta(2 \sin p_i - 2 \lambda_\alpha), \\
  2\pi J_\alpha &= \sum_{\beta=1}^{M} \theta(\lambda_\alpha - \lambda_\beta) - \sum_{i=1}^{N} \theta(2\lambda_\alpha - 2 \sin p_i), 
           \label{eq:bethe-ansatz-equations-fermions}
\end{align}
where $i=1,\dotsc,N$, $\alpha = 1,\dotsc,M$, 
\begin{align}
  \theta(p) &= -2 \tan^{-1}\left(\frac{2pt}{U}\right) \in [-\pi, \pi),
\end{align}
and $I_i, J_\alpha$ are specific integers (or half-odd integers) for $N,M$ even (or odd) that uniquely label the energy eigenstate.  The energy eigenvalue is then given by $E = 2t \sum_{i} \left(1- \cos(p_i)\right)$. For the ground state, we must choose $I_i = -\frac12(N+1) + i$ and $J_\alpha = -\frac12 (M+1) + \alpha$. The ground state for the attractive Hubbard model can be obtained from the repulsive case using the relation
\begin{align}
      E_0(N_\uparrow, N_\downarrow, U) &= E_0(N_\uparrow, L-N_\downarrow, -U) \nonumber \\
      &\quad + UN_\uparrow - 2 t_\downarrow (L - 2 N_\downarrow).
\end{align}
The exact ground-state energy obtained by this procedure for $N_\ua=N_\da=5$ is shown in the left plot of \cref{fig:bethe-ansatz} as a dashed line labeled ``Bethe ansatz.'' 
We also show results obtained from perturbation theory up to second order. 
We note that our method, labeled as ``Worm algorithm,'' is able to precisely reproduce the exact results from the Bethe ansatz once we extrapolate to the continuous time limit $\varepsilon \to 0$ even at $\beta=100$.  
For comparison, we also plot the CL results from Ref.~\cite{PhysRevD.96.094506} on the same plot. 
We notice that CL reproduces the results quite well for small values of the couplings, but starts to deviate when $\gamma \gtrsim 2.5$. 
To confirm that the worm algorithm reproduces the exact Bethe ansatz calculations in the strong coupling regime, where perturbation theory clearly breaks down, we extend these results to higher values of $\gamma$ in the right plot of \cref{fig:bethe-ansatz}. 

There is a small difference between the exact results from Bethe ansatz in the continuum as compared to the lattice, as shown in the left plot of \cref{fig:bethe-ansatz-beta-extrapolation}. In the right plot, we show that the worm algorithm is able to resolve this difference.
the exact continuum and lattice Bethe ansatz results (at $\beta\EnergyFermi=\infty$).  
We also show the worm algorithm results at different values of $\beta\EnergyFermi$, which indeed converge to the correct lattice answer in the limit $\beta \EnergyFermi \to \infty$.

\begin{figure}[htbp]
\centering
  \includegraphics[width=\FigureWidth]{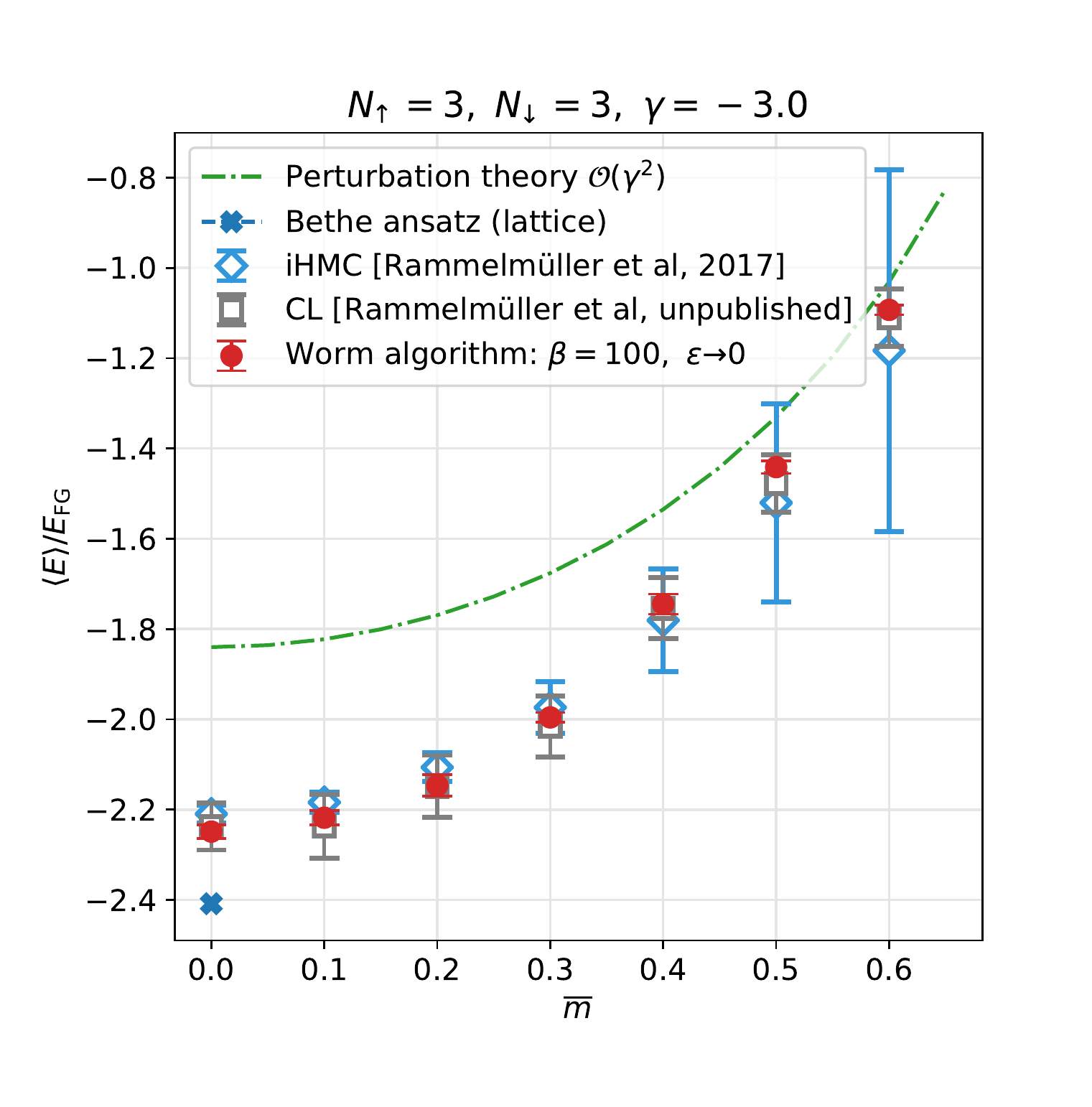}
  \caption{Energy as a function of mass-imbalance $\mbar$. Note that the lattice model is invariant under $\mbar\to -\mbar$, which implies that the curve is quadratic in $\mbar$ for small values of $\mbar$. We show our data using the worm algorithm along with the iHMC results of Ref.~\cite{PhysRevD.96.094506}, unpublished CL data given to us by the authors of Ref.~\cite{PhysRevD.96.094506}, and second order perturbation theory. The apparent discrepancy at $\mbar=0$ between our results and the results from the Bethe ansatz is addressed in \cref{fig:33-attractive-mbar-0-beta-ex}.}
  \label{fig:33-attractive-mbar-vary}
\end{figure}

\subsection{Mass-Imbalanced Systems}

In contrast to the mass-balanced case, no exact solution is known for the general mass-imbalanced case ($\mbar \neq 0$).
However, our algorithm is again very efficient even in such cases. In this section, we compute some spin-balanced results and check our results against second-order perturbation theory and find good agreement for small values of $\gamma$.  
\Cref{fig:55-repulsive-mbar=0.8} shows our results for the ground state energy of $5+5$ particles as a function of the coupling for a high mass-imbalance of $\mbar=0.8$. We perform computations at $\beta=100$, which we find to be sufficient to make our point.

As can be seen from the left plot of \cref{fig:55-repulsive-mbar=0.8}, our results agree very well with second-order perturbation theory for up to $\gamma \sim 3.0$, while the CL results disagree with both perturbation theory and our method when $\gamma \gtrsim2$. In the right plot of \cref{fig:55-repulsive-mbar=0.8}, we extend this to the regime of very strong repulsion. It was suggested in Ref.~\cite{PhysRevD.96.094506} that the flattening of the ground-state energy as a function of $\gamma$ for strong repulsive couplings, obtained using the CL method, could be a physical effect. 
However,  the disagreement with exact Bethe ansatz calculations at $\mbar=0$ (\cref{fig:bethe-ansatz}) and with the worm algorithm at a high mass-imbalance of $\mbar=0.8$ (\cref{fig:55-repulsive-mbar=0.8}) shows that the observed flattening is an artifact of the CL method.

\begin{figure*}
  \includegraphics[width=\FigureWidth,valign=t]{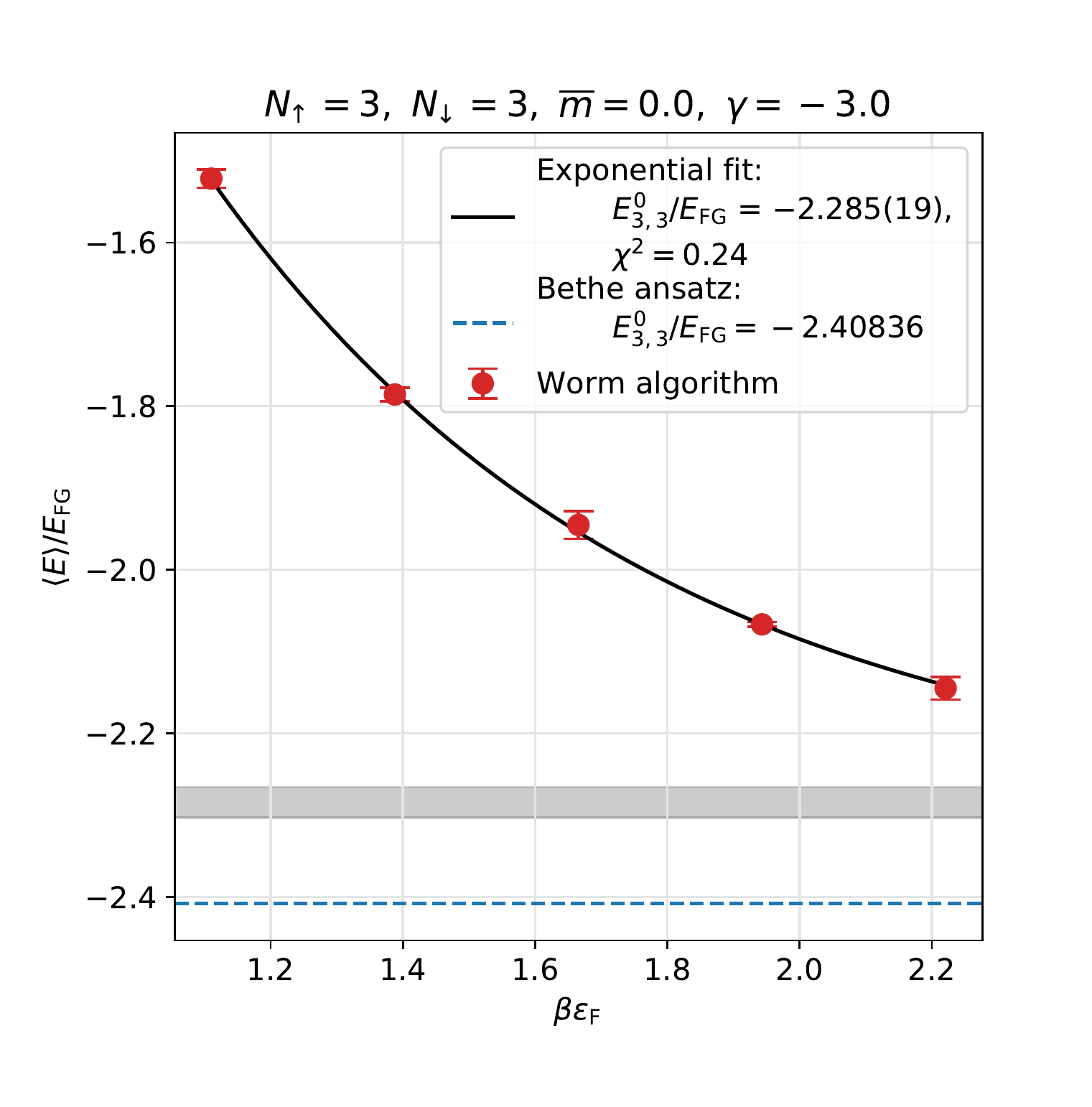}\hfill
  \includegraphics[width=\FigureWidth,valign=t]{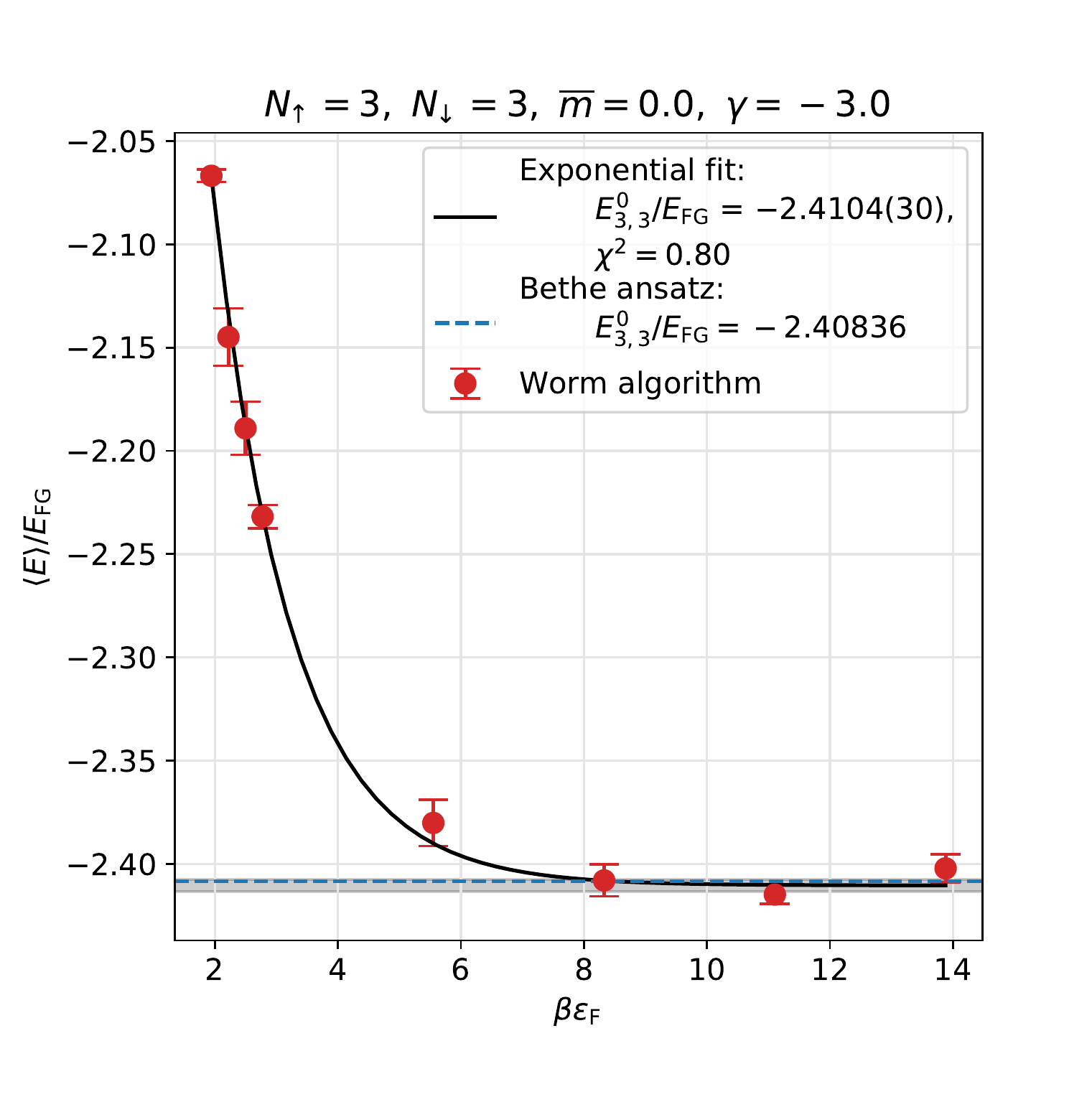}
  \caption{The two plots shown above compare the systematic errors in extracting the ground state energy $E^0_{N_\ua,N_\da}$ due to the fitting range in $\beta$ used to fit the data to \cref{eq:Energyfit}. The data is from the mass-balanced case ($\mbar = 0$) with $N_\ua = N_\da = 3$ particles with $\gamma = -3.0$ and $\LX = 40$ (see  \cref{fig:33-attractive-mbar-vary}). The left plot shows the fit for the range $25\leq \beta \leq 75$, while the plot on the right is for the fit in the range $100\leq \beta \leq 500$. The extracted ground state energy along with the error is shown as a shaded region. The exact answer from the Bethe ansatz is shown at a dotted line in the both the plots.
  \label{fig:33-attractive-mbar-0-beta-ex}}
\end{figure*}

In \cref{fig:33-attractive-mbar-vary}, we perform a comparison with the results of Ref.~\cite{PhysRevD.96.094506} 
for $N_\ua=N_\da=3$ fermions with a fixed attractive coupling $\gamma=-3.0$ across a range of mass-imbalances. 
We observe that imaginary-mass Hybrid Monte Carlo (iHMC) has large errors for higher values of $\mbar$, but our method performs consistently across a wide range of mass-imbalances. 
We also show unpublished data from the CL approach, shared with us by the authors of Ref.~\cite{PhysRevD.96.094506}. These CL results seem to converge to the correct values, in contrast to the repulsive case discussed earlier.
The prediction from second order perturbation theory is also plotted. At $\mbar=0$, the exact Bethe ansatz result clearly disagrees with our algorithm. However, our results are obtained at $\beta=100$ and the exact answer is entirely within the $\Order(1/\beta)$ error expected from higher excited states.  We discuss this issue in more detail in the next section.

\section{Extracting the ground state energy}
\label{sec:extractGSE}

In the above section, we presented results for $\langle E\rangle$ at a fixed value of $\beta=100$. We found in \cref{fig:bethe-ansatz} that the results almost agreed with the exact ground state. The small disagreement was shown to be an effect of $\beta$ not being large enough in \cref{fig:bethe-ansatz-beta-extrapolation}. This suggests that $\beta=100$ is not guaranteed to be large in all the cases. In fact, in \cref{fig:33-attractive-mbar-vary} we found that our value for the energy obtained at $\beta=100$ disagreed with the exact ground state energy from the Bethe ansatz at $\mbar=0$ by several standard deviations. This shows that it is important to be able to perform a systematic extrapolation to the $\beta \rightarrow \infty$ limit to extract the ground state energy.

At $\mbar=0.0$, \cref{fig:33-attractive-mbar-vary} shows a deviation of roughly $0.009$ between the exact energy and the energy at $\beta=100$ in bare units (with $\EnergyFG = 0.056$). While this is within the expected $\Order(1/\beta)$ corrections, it is important to be able to perform a systematic extrapolation to $\beta\to\infty$ to extract the ground-state energy. The traditional procedure followed in the literature (which we refer to as Method I) is to compute $\langle E\rangle$ at several values of $\beta$ and then fit to the form
\begin{align}
    \langle E \rangle \ =\ E^0_{N_\ua,N_\da} + A \exp(-B \beta).
    \label{eq:Energyfit}
\end{align}
While this approach is surely reasonable at sufficiently large values of $\beta$, it is a priori not clear what range of $\beta$ should be chosen for the fit.  We believe a common misconception is that the correct range of $\beta$ can be determined by increasing the upper limit of $\beta$ until the
above form begins to fit the data well in a range.  This of course depends on the precision to which the average energies are computed.  Here we show that even if the errors are in the one-percent range, which is usually difficult in many cases, we can get a good fit but with wrong results if we do not choose a sufficiently large range of $\beta$.

\begin{figure*}[htbp]
  \newlength\FigureWidthBig
  \setlength\FigureWidthBig{2.5in}
  \def\FigScaleFactor{1.00}
\includegraphics[width=\FigScaleFactor\textwidth,height=\FigScaleFactor\textheight,keepaspectratio]{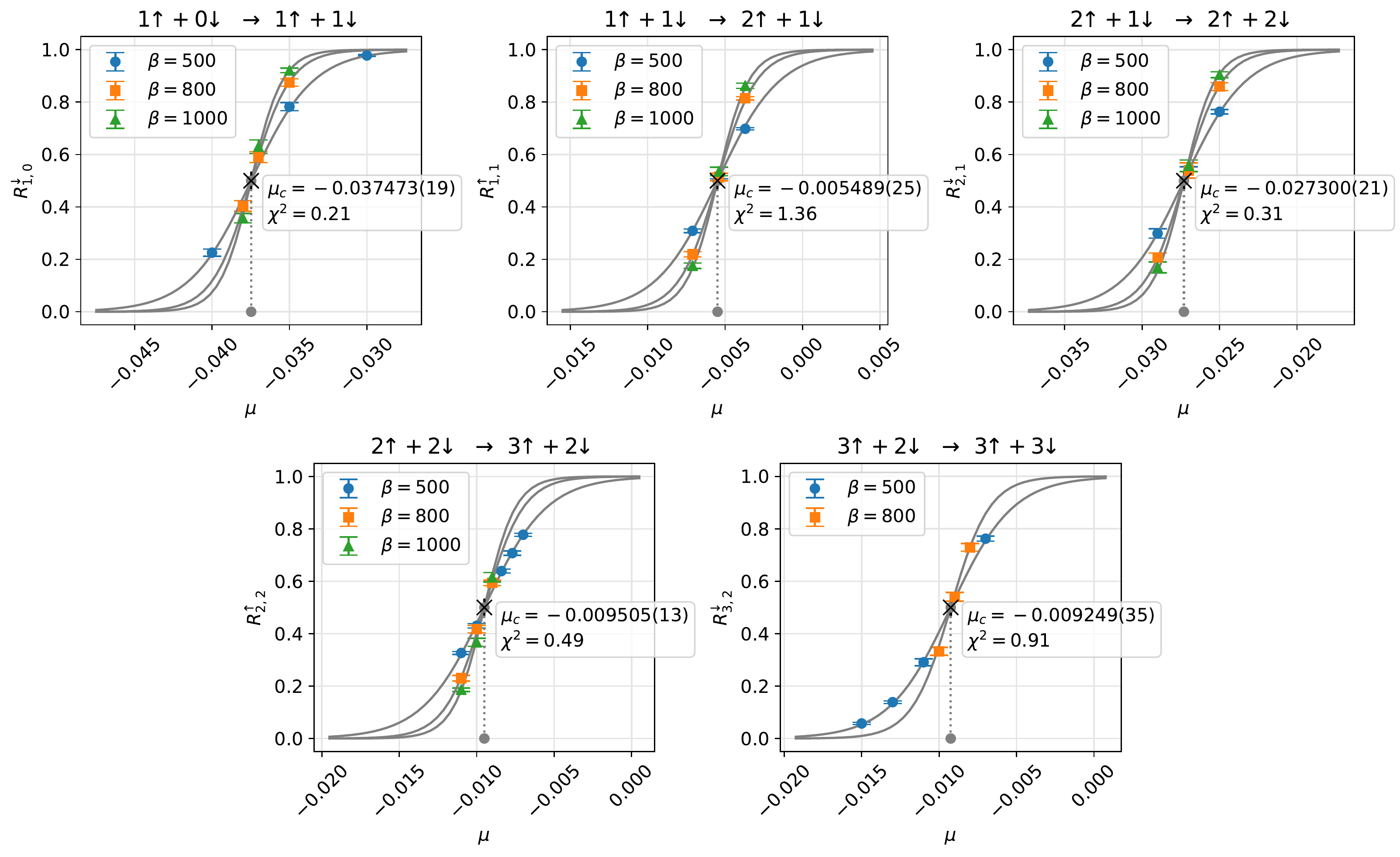}
  \caption{Extraction of the ground state energy by the method of adding one particle at a time and fitting to the one parameter function \cref{eq:fitform}. The calculations shown here are for the mass-imbalanced case with $\mbar=0.5$, $\gamma=-3.0$, and $\varepsilon=0.01$.
  }
  \label{fig:33-precise-method}
\end{figure*}

\begin{figure*}[hbt]
  \centering
  \includegraphics[width=\FigureWidth]{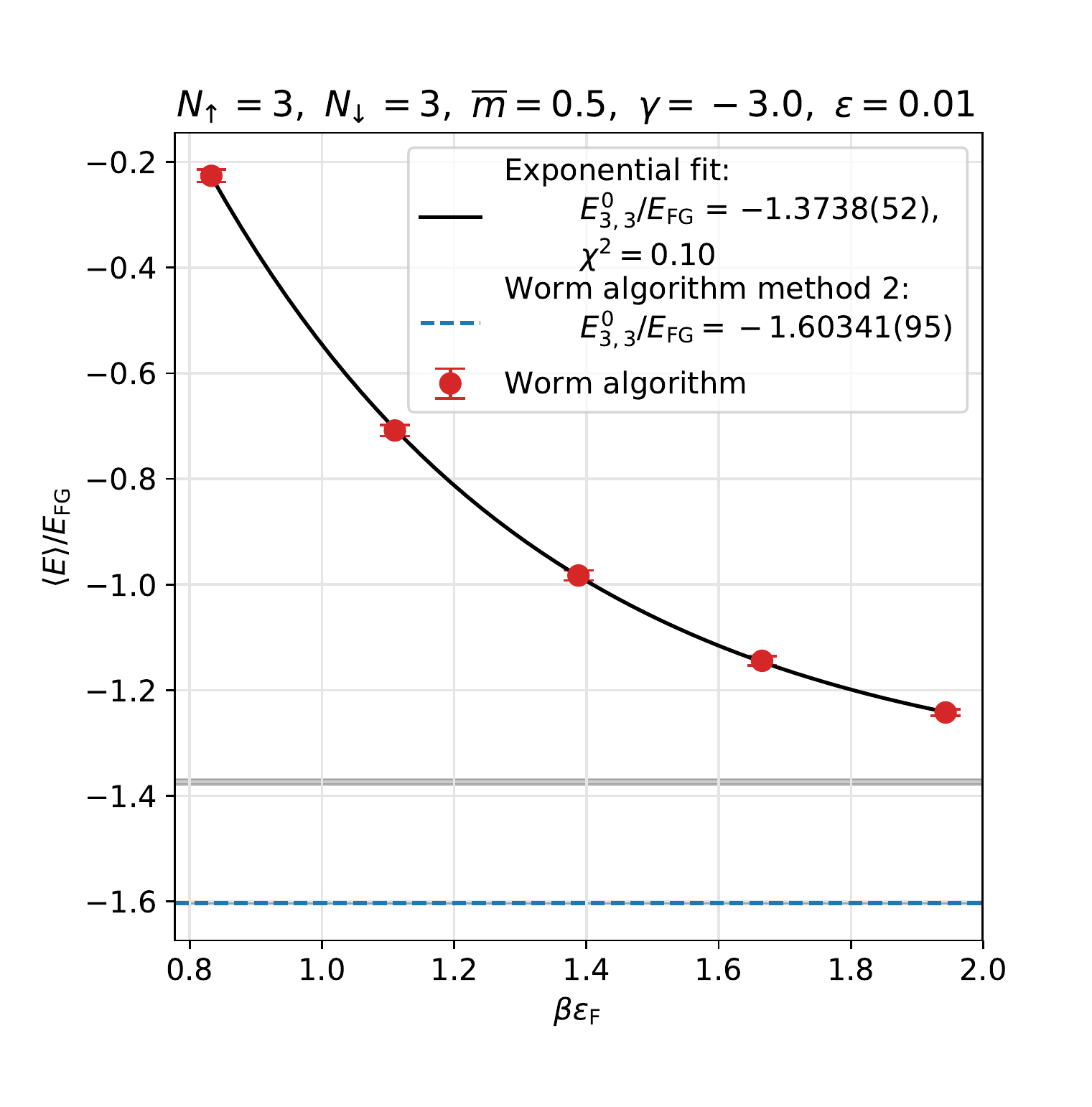}
  \includegraphics[width=\FigureWidth]{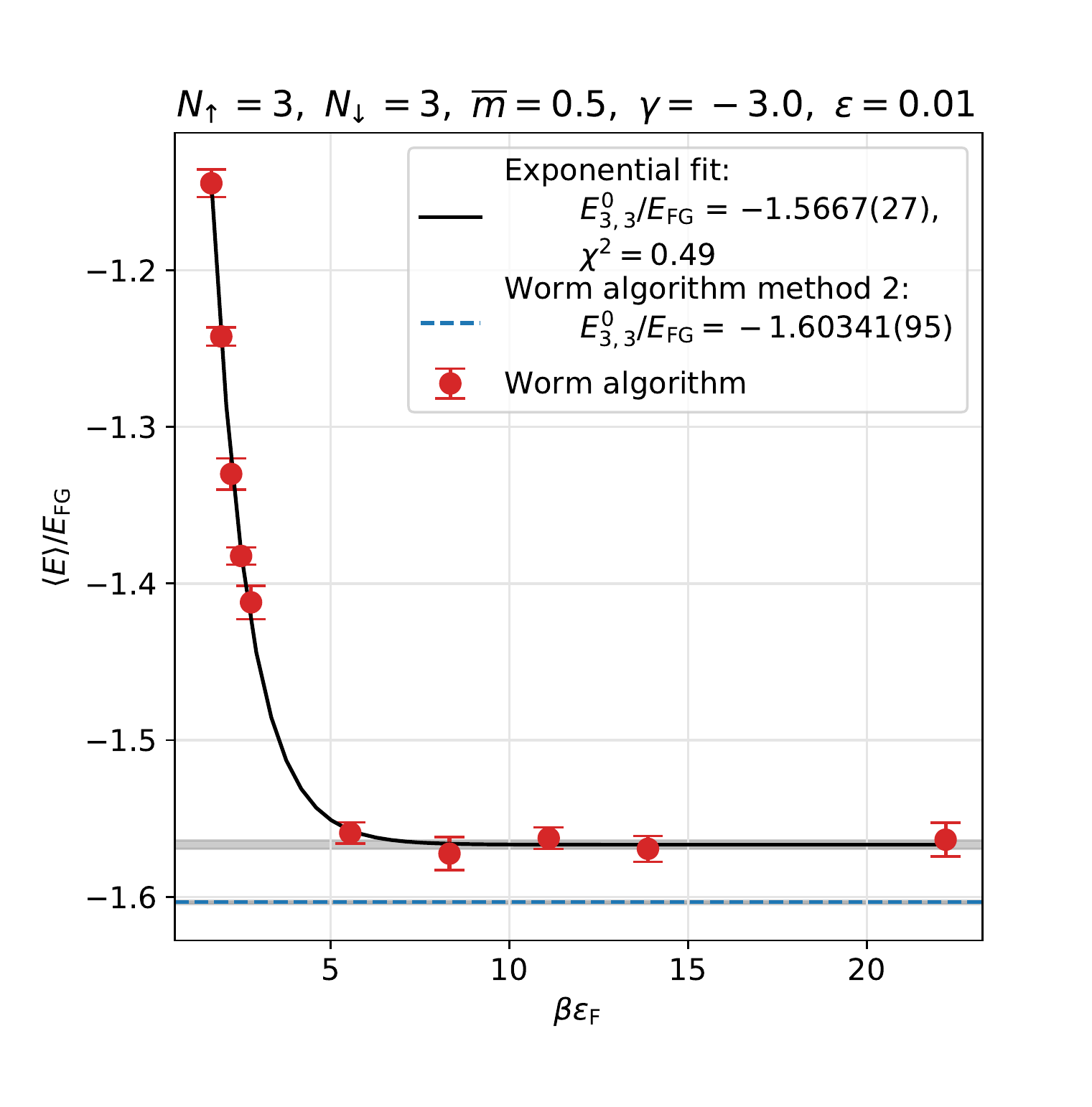}\hfill
  \caption{Comparison of errors in determining the ground state energy by the two methods discussed in the text. The data and fits shown are for the traditional method (method I) in two different ranges of $\beta$. The flat solid line is the result for the ground state energy obtained from the fit. The dashed line is the result for the ground state energy obtained from method II (see \cref{fig:33-precise-method}). The results from method the two methods don't have to agree since the calculations are performed at a non-zero value of $\varepsilon$.}
  \label{fig:33-precise-method-comparison}
\end{figure*}

To demonstrate the problem, in \cref{fig:33-attractive-mbar-0-beta-ex} we show the fit for two ranges of $\beta$ at $\gamma = -3.0$ in the mass-symmetric case where we know the exact answer, shown as the dashed line. As can be seen from the figure, the fit in the range $\beta \in [40,80]$ ($\beta\EnergyFermi \in [1.11, 2.22]$) is excellent but gives us  $E^{0}_{N_\ua,N_\da}/\EnergyFG = -2.285(19)$, which is different from the exact result of $E^{0}_{N_\ua,N_\da}/\EnergyFG = -2.40836$
by several standard deviations. On the other hand notice that the fit in the range $\beta \in [70, 500]$ ($\beta \EnergyFermi \in [1.94, 13.88]$) gives a better answer of 
$E^{0}_{N_\ua,N_\da}/\EnergyFG = -2.4104(30)$. This also explains the deviation in \cref{fig:33-attractive-mbar-vary} at a fixed value of $\beta=100$ noted in the previous paragraph. 

These observations suggest a need for a complementary way to compute $E^0_{N_\ua,N_\da}$, which can be compared with the above method. 
Below we discuss one such method (which we refer to as Method II) which we believe gives better precision although the analysis is more involved. Since the worm algorithm allows us to study a variety of particle number sectors efficiently, we can efficiently build the desired particle-number sector by adding one particle at a time. 
Consider a situation where we can tune the chemical potentials $\mu_\sigma$ close to a critical value so that the average particle numbers fluctuate between $(N_\ua,N_\da)$ and $(N_\ua+1,N_\da+1)$. Near such a critical point, we can develop efficient worm algorithms to sample configurations where $N_\ua$ fluctuates by one while $N_\da$ remains fixed. We can then accurately measure the ratios like
\newcommand\ZMu[1]{Z_{\mu}^{#1}}
\begin{align}
    R^\ua_{N_\ua,N_\da} = \frac{\ZMu{N_\ua+1,N_\da}}{\ZMu{N_\ua,N_\da}+\ZMu{N_\ua+1,N_\da}},
\end{align}
where $\ZMu{\NN}$ as the partition function defined in \cref{eq:Z-mu} restricted to a fixed particle number sector,
\begin{align}
    \ZMu{N_\ua, N_\da} &= 
    \Trace\Big( {e}^{-\beta H_\mu }\Big)\Big|_{\NN}.
    \label{zratio}
\end{align}
Assuming $\beta$ is sufficiently large so that only the ground states contribute, we must have
\begin{align}
    R^\ua_{N_\ua,N_\da} = 
    \frac{g \exp(-\beta (\mu_c-\mu_\ua))}{1+g\exp(-\beta (\mu_c-\mu_\ua))}
    \label{eq:fitform}
\end{align}
where
\begin{align}
    \mu_c = E^0_{N_\ua+1,N_\da}-E^0_{N_\ua,N_\da}
\end{align}
is the difference in the ground-state energies of the two particle number sectors, and $g$ is the ratio of their degeneracies, which is a fraction typically made up of small integers.  If $g$ can be determined from the knowledge that it is a fraction of small integers, the fit function \eqref{eq:fitform} has a single free parameter ($\mu_c$) for all values of $\beta$ (sufficiently large) and $\mu_\ua$. 
This fitting procedure gives very precise values for the critical chemical potential $\mu_c$. A similar procedure can be adapted to compute the difference $E^0_{N_\ua,N_\da+1}-E^0_{N_\ua,N_\da}$. Absolute energies can be computed by adding such differences.

In order to demonstrate that method II gives us more accurate answers as compared to method I and more importantly does not depend on the range of $\beta$ used in the analysis we have extracted the ground state using both the methods at $\mbar = 0.5$ and $\gamma = -0.3$. However, for this study we limited our effort to a fixed non-zero value of $\varepsilon = 0.01$. It should be noted that when $\varepsilon \neq 0$ the definition of the ground state energy obtained from the two methods can disagree due to $\Order(\varepsilon)$ errors. But it can still give us a sense of the magnitude of the errors in computation. The extrapolation to $\varepsilon \rightarrow 0$ requires further effort but can be done if necessary.

In \cref{fig:33-precise-method} we show our results for method II. We begin with the system containing a single particle ($N_\ua=1,N_\da = 0$) whose ground state energy is known to be zero. We then add particles slowly to reach $N_\ua=N_\da=3$ in five steps. The results from each step are shown in the figure by plotting the ratio $R^\sigma_{N_\ua,N_\da}$ as a function of $\mu$. The solid lines in each plot are a combined fit to the form \cref{eq:fitform} with one parameter $\mu_c$. This gives us the following results:
\begin{align*}
E^0_{1, 1} - E^0_{1, 0} &= -0.037473(19),\\
E^0_{2, 1} - E^0_{1, 1} &= -0.005489(25),\\
E^0_{2, 2} - E^0_{2, 1} &= -0.027300(21),\\
E^0_{3, 2} - E^0_{2, 2} &= -0.009505(13),\\
E^0_{3, 3} - E^0_{3, 2} &= -0.009249(35).
\end{align*}
Adding the all the values of $\mu_c$ we obtain $E^0_{3,3} = -0.089016(53)$ which gives $E^0_{3,3}/\EnergyFG = -1.60341(95)$.

To compare the errors obtained from method II with method I we compute  $\langle E\rangle$ at several values of $\beta$. In \cref{fig:33-precise-method-comparison} we show fits of this data to the form \cref{eq:Energyfit} in two different ranges of $\beta$. Again we see that in method I, the range of $\beta$ at larger values gives a slightly different estimate of the ground-state energy as compared to the range at smaller values. Due to time discretization errors that are present at $\varepsilon = 0.01$, the estimate of the ground state energy computed using the two methods can disagree. Here we focus on the error in this estimate and observe that it is a factor of three more in method I as compared to method II, although individual data points were all obtained with the same precision of roughly one percent.

\section{Higher Dimensions}
\label{sec:higher-dimensions}

The quantum Monte Carlo method we have developed in this work is guaranteed to work well only for systems with hard-core bosons and not fermions. In one dimension, where fermions are identical to hard-core bosons in many cases, our method can be used to study fermionic systems as well. We have demonstrated this in \cref{sec:1d-results}. 
However, as we pointed out in \cref{sec:sign}, we can also study a system of fermions in higher dimensions if we can compute the fermion sign $\langle S\rangle$ accurately. We believe this may be possible in the context of few-body physics by combining ideas of fermion bags proposed recently \cite{PhysRevD.96.114502} with ideas of exponential error reduction proposed several years ago \cite{Luscher:2001up}. While a complete study of this more difficult problem is a topic for another paper, here we provide some evidence that our method can indeed be used for computing the ground-state energy even in higher dimensions with fermions.
For this purpose, we compute the ground state energy of our Hamiltonian \eqref{eq:lattice-model} in three spatial dimensions with $\LX=8$  lattice for the mass-balanced system ($m_\ua=m_\da=1$) with $N_\ua = N_\da=2$ at $U=-3.9570$, which corresponds to the unitary fixed point in the continuum limit. The exact ground-state energy for this system was computed some years ago as a benchmark calculation by an explicit diagonalization of the Hamiltonian using the Lanczos algorithm and was found to be $E^{0f}_{2,2} \approx 0.1042$ \cite{PhysRevA.83.063619}. Here we reproduce this result using our approach.

\begin{figure}[htbp]
  \centering
  \includegraphics[width=\linewidth]{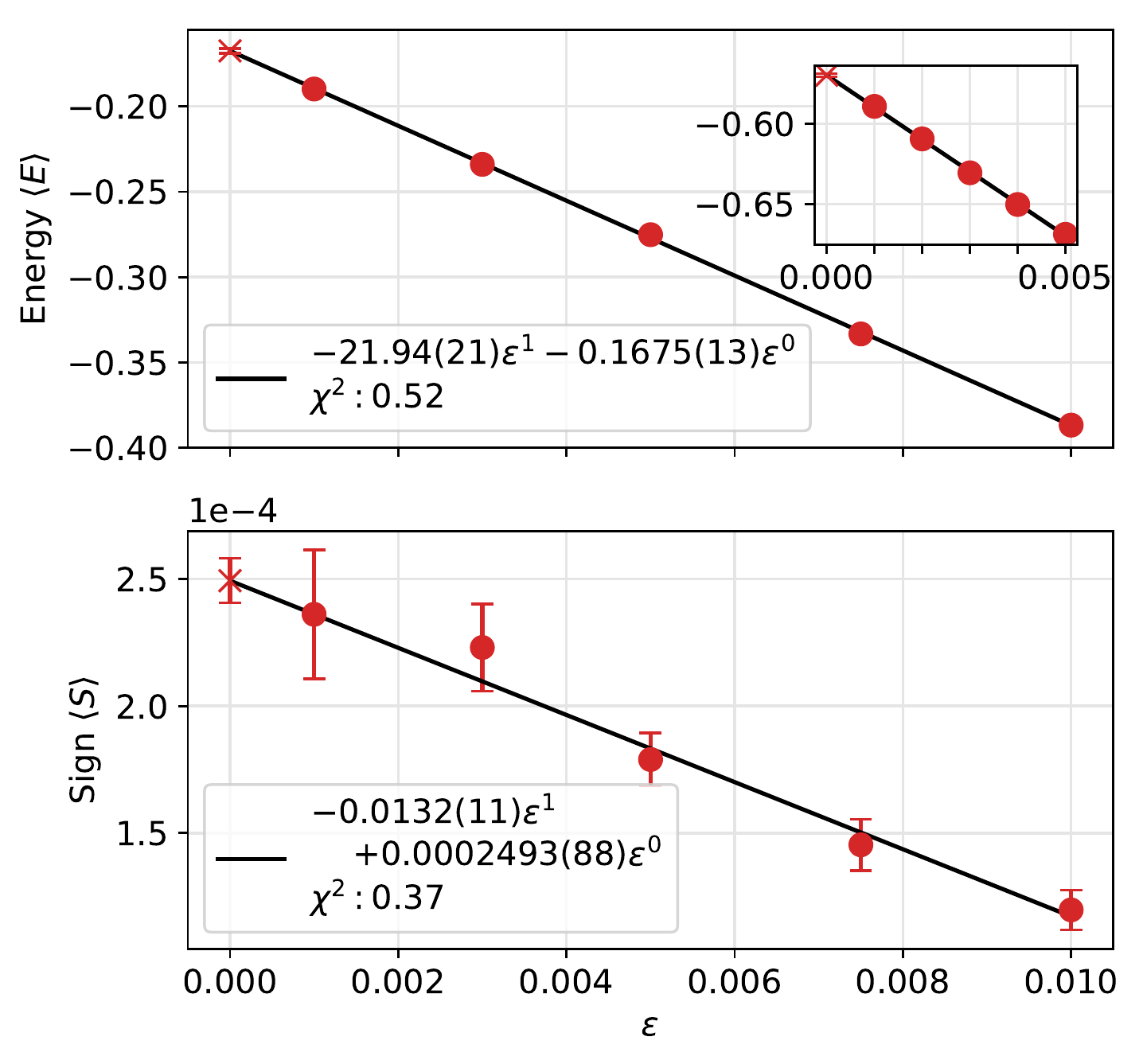}
  \caption{Plots of the bosonic energy $\langle E\rangle$ and the fermion sign $\langle S\rangle$ as a function of the temporal lattice spacing in the three dimensional mass-balanced model ($m_\ua=m_\da=1$) at $\LX=8$, $U=-3.9570$ for $N_\ua=N_\da=2$ particles. The inset shows bosonic energy as a function of $\varepsilon$ at $\LX=4$. The solid lines are linear fits to extract the continuum time limit. }
  \label{fig:highdim}
\end{figure}

We first performed calculations at several values of $\beta$ ranging from $10$ to $50$ at $\varepsilon=0.01$. The results for the bosonic energy $\langle E\rangle$ are shown  in \cref{tab:betavsbe}. Since the bosonic energy does not change much between $\beta=30$ and $50$ we assume that $\beta=30$ is sufficiently large and perform a careful extrapolation of the $\varepsilon$ errors there. In \cref{fig:highdim} we show our results for this extrapolation for both bosonic energy $\langle E\rangle$ and the fermion sign $\langle S\rangle$. Note that errors due to a non-zero value of $\varepsilon$ are linear at leading order and hence can be large. The solid lines in the figure are fits that we used to extract $\varepsilon \rightarrow 0$ limit. In this time continuum limit we find that $\langle E\rangle = -0.1675(20)$ and $\langle S\rangle = 0.00025(2)$. Using \cref{eq:fermigse} we estimate $E^{0f}_{2,2} \approx 0.1090(40)$, which is in reasonable agreement with the exact result $E^{0f}_{2,2} = 0.1042$ \cite{PhysRevA.83.063619}.

\begin{table}[hb]
  \SetTableProperties{}
  \begin{tabular}{c  *{5}{S[table-parse-only]}}
    \TopRule
    $\beta$ & 10 & 15 & 20 & 30 & 50 \\
    $\< E \>$ & 
    -0.12(1) & -0.322(5) & -0.368(5) & -0.387(2) & -0.387(2) \\
   \BotRule
  \end{tabular}
  \caption{$\langle E\rangle$ for various values of $\beta$ in the three dimensional mass-balanced model ($m_\ua=m_\da=1$) at $\LX=8$, $U=-3.9570$ and $\varepsilon = 0.01$ for $N_\ua=N_\da=2$ particles.
  \label{tab:betavsbe}}
\end{table}

\section{Conclusions}
\label{sec:conclusions}

In this work, we proposed a worldline based approach to few-body physics where fermions are formulated as hard-core bosons, since they incorporate one of the ingredients of the Pauli exclusion principle, which is that two identical fermions cannot exist at the same spacetime point. On the other hand, hard-core bosons do not capture the fermion permutation sign, which needs to be taken into account explicitly before our method can truly be applicable to fermionic systems. Fortunately, in one spatial dimension, fermion permutations can only occur over the boundaries and the fermion permutation sign is positive in many cases. 

Our approach is complementary to the well-developed AFQMC methods for fermionic systems, which unfortunately suffer from sign problems even in one spatial dimension. To demonstrate the power of our method, we showed that we can reproduce some exact results for the one-dimensional Hubbard model, obtained using the Bethe ansatz, for a wide range of couplings in the mass-balanced case. Our method can easily be applied to the mass-imbalanced case where a general exact solution is not known. We used our approach to show that the results from the CL method, recently presented in Ref.~\cite{PhysRevD.96.094506}, yields  wrong values for repulsive interactions. This must be related to the `fat-tailed'  distributions of the observables in the CL method, as noted in the appendix of Ref.~\cite{PhysRevD.96.094506}. On the other hand, based on the data shared with us by the authors of that paper, the CL method seems more robust on the attractive side in the parameter range studied.

Extending our approach to higher dimensions is straightforward, although we have to confront the fermion sign problem which is equally severe for all types of interactions. Unlike the AFQMC methods, there is no particular advantage for attractive interactions as compared to repulsive interactions, although we do see that the sign problem becomes slightly milder in the attractive case. We presented some evidence that, at least for few-body physics, we may be able tame the sign problem using ideas of fermion bags and exponential error reduction. In particular, we were able to reproduce a benchmark calculation done a few years ago with four fermions at unitarity \cite{PhysRevA.83.063619}.

Finally, unlike the AFQMC method, our worldline approach can be formulated directly in the time continuum limit \cite{PhysRevLett.77.5130,PhysRevB.72.035122,PhysRevD.96.114502}, although in this work we did not exploit this advantage. Instead, we studied the time discretization errors. We found that they are linear in the temporal lattice spacing $\varepsilon$ and can be eliminated by a simple extrapolation. Without such an extrapolation we would not have been able to compare our results with the exact results from the Bethe ansatz.

\section*{Acknowledgments}

We would like to thank Jens Braun, Joaqu\'in Drut and Lukas Rammelm\"uller for sharing their data published in Ref.~\cite{PhysRevD.96.094506}, the unpublished data that appears in \cref{fig:33-attractive-mbar-vary}, and for productive discussions about their results.
We would also like to thank Dean Lee, Abhishek Mohapatra and Roxanne Springer for helpful conversations about effective field theories. 
The material presented here is based upon work supported by the U.S. Department of Energy, Office of Science, Nuclear Physics program under Award Numbers DE-FG02-05ER41368. 

\appendix

\begin{table*}
  \SetTableProperties{}
    \begin{tabular}{*{5}{c} *{3}{S[]} }
\TopRule
{$\mu_\ua$} & {$\mu_\da$} & {$m_\ua$} & {$m_\da$} & {$U$} 
&{ 
\parbox{2cm}{\vspace{0.5em}$\< E \>$ $\begin{array}{c} \text{MC} \\[-0.5em] \text{Exact} \end{array}$ \vspace{0.5em}} }
&{ 
\parbox{2cm}{\vspace{0.5em}$\< N_\ua \>$ $\begin{array}{c} \text{MC} \\[-0.5em] \text{Exact} \end{array}$ \vspace{0.5em}} }
&{ 
\parbox{2cm}{\vspace{0.5em}$\< N_\da \>$ $\begin{array}{c} \text{MC} \\[-0.5em] \text{Exact} \end{array}$ \vspace{0.5em}}} \\
\HeaderRule
%--------------------------------------------------------------------------------
\multirow{2}{*}{\tablenum{1.00}}	&	\multirow{2}{*}{\tablenum{-2.00}}	&	\multirow{2}{*}{\tablenum{1.50}}	&	\multirow{2}{*}{\tablenum{0.50}}	&	\multirow{2}{*}{\tablenum{-4.00}}	&	-0.552802(53)	&	1.415660(25)	&	0.572415(24)\\
	&		&		&		&		&	-0.552791	&	1.415684	&	0.572407\\

\multirow{2}{*}{\tablenum{2.00}}	&	\multirow{2}{*}{\tablenum{0.00}}	&	\multirow{2}{*}{\tablenum{1.50}}	&	\multirow{2}{*}{\tablenum{0.50}}	&	\multirow{2}{*}{\tablenum{-2.00}}	&	0.674356(26)	&	1.706027(29)	&	0.732806(31)\\
	&		&		&		&		&	0.674351	&	1.706015	&	0.732805\\

\multirow{2}{*}{\tablenum{1.00}}	&	\multirow{2}{*}{\tablenum{4.00}}	&	\multirow{2}{*}{\tablenum{1.50}}	&	\multirow{2}{*}{\tablenum{0.50}}	&	\multirow{2}{*}{\tablenum{0.00}}	&	3.525805(51)	&	1.135221(19)	&	1.605845(16)\\
	&		&		&		&		&	3.525821	&	1.135242	&	1.605851\\

\multirow{2}{*}{\tablenum{2.50}}	&	\multirow{2}{*}{\tablenum{2.00}}	&	\multirow{2}{*}{\tablenum{1.50}}	&	\multirow{2}{*}{\tablenum{0.50}}	&	\multirow{2}{*}{\tablenum{2.00}}	&	1.739032(49)	&	1.574851(25)	&	0.333418(20)\\
	&		&		&		&		&	1.738968	&	1.574855	&	0.333407\\

\multirow{2}{*}{\tablenum{1.00}}	&	\multirow{2}{*}{\tablenum{4.00}}	&	\multirow{2}{*}{\tablenum{1.50}}	&	\multirow{2}{*}{\tablenum{0.50}}	&	\multirow{2}{*}{\tablenum{4.00}}	&	2.75805(10)	&	0.357677(24)	&	1.354424(28)\\
	&		&		&		&		&	2.758091	&	0.357684	&	1.354429\\
%--------------------------------------------------------------------------------
\BotRule
  \end{tabular}
  \caption{Comparison of our Monte Carlo method with the exact results for a $2\times 2$ space-time lattice in $1+1$ dimensions. We show results for the the average energy $\< E \> $ and average particle numbers $N_\ua, N_\da$.  }
  \label{tab:exact-2x2}
\end{table*}

\newcommand\TableSectionSeperator{\MidRule}
\colorlet{highlightrow}{black!8!white}

\begin{table*}
    \SetTableProperties{}
   \begin{tabular}{*{8}{c} c *{4}{S[table-format=4.6(2)]}}
\TopRule
%-------------------------------------%
{\multirow{2}{*}{$d$}} & {\multirow{2}{*}{$L_X$}} & {\multirow{2}{*}{$\beta$}} & {\multirow{2}{*}{$\mu_\ua$}} & {\multirow{2}{*}{$\mu_\da$}} & {\multirow{2}{*}{$m_\ua$}} & {\multirow{2}{*}{$m_\da$}} & {\multirow{2}{*}{$U$}} &   & \multicolumn{3}{c}{Worm Algorithm} & {\multirow{2}{*}{Exact}}\\
  &   &   &   &   &   &   &   &   & \multicolumn{1}{c}{$\varepsilon=0.001$} & \multicolumn{1}{c}{$\varepsilon=0.0005$} & \multicolumn{1}{c}{$\varepsilon \to 0$}\\
\HeaderRule
%-------------------------------------%
 &  &  &  &  &  &  &  & {$\< E \> $} & 8.29402(69) & 8.29095(69) & 8.2875(17) & 8.2881\\
 &  &  &  &  &  &  &  & {$ \< N_\ua \>$} & 2.84525(34) & 2.84502(91) & 2.84532(81) & 2.8454\\
\multirow{-3}{*}{\tablenum{1}} & \multirow{-3}{*}{\tablenum{6}} & \multirow{-3}{*}{\tablenum{10}} & \multirow{-3}{*}{\tablenum{2.00}} & \multirow{-3}{*}{\tablenum{4.00}} & \multirow{-3}{*}{\tablenum{1.50}} & \multirow{-3}{*}{\tablenum{0.50}} & \multirow{-3}{*}{\tablenum{2.00}} & {$ \< N_\da \>$} & 3.95296(18) & 3.95307(31) & 3.95264(32) & 3.9530\\
\TableSectionSeperator
 &  &  &  &  &  &  &  & {$\< E \> $} & 3.67766(86) & 3.6726(23) & 3.6663(22) & 3.6668\\
 &  &  &  &  &  &  &  & {$ \< N_\ua \>$} & 3.26671(48) & 3.26674(48) & 3.2659(13) & 3.2644\\
\multirow{-3}{*}{\tablenum{1}} & \multirow{-3}{*}{\tablenum{6}} & \multirow{-3}{*}{\tablenum{10}} & \multirow{-3}{*}{\tablenum{-1.00}} & \multirow{-3}{*}{\tablenum{3.00}} & \multirow{-3}{*}{\tablenum{1.50}} & \multirow{-3}{*}{\tablenum{0.50}} & \multirow{-3}{*}{\tablenum{-2.00}} & {$ \< N_\da \>$} & 5.15396(40) & 5.15293(40) & 5.15134(86) & 5.1509\\
\TableSectionSeperator
 &  &  &  &  &  &  &  & {$\< E \> $} & 0.09151(23) & 0.09179(33) & 0.09299(83) & 0.0928\\
 &  &  &  &  &  &  &  & {$ \< N_\ua \>$} & 0.79921(42) & 0.79698(24) & 0.79464(19) & 0.7953\\
\multirow{-3}{*}{\tablenum{2}} & \multirow{-3}{*}{\tablenum{2}} & \multirow{-3}{*}{\tablenum{10}} & \multirow{-3}{*}{\tablenum{0.40}} & \multirow{-3}{*}{\tablenum{0.30}} & \multirow{-3}{*}{\tablenum{1.50}} & \multirow{-3}{*}{\tablenum{0.50}} & \multirow{-3}{*}{\tablenum{2.00}} & {$ \< N_\da \>$} & 0.40562(38) & 0.41007(26) & 0.41527(76) & 0.4148\\
\TableSectionSeperator
 &  &  &  &  &  &  &  & {$\< E \> $} & 5.2475(53) & 5.2571(37) & 5.2661(76) & 5.2631\\
 &  &  &  &  &  &  &  & {$ \< N_\ua \>$} & 3.56841(94) & 3.57136(67) & 3.5741(13) & 3.5732\\
\multirow{-3}{*}{\tablenum{2}} & \multirow{-3}{*}{\tablenum{2}} & \multirow{-3}{*}{\tablenum{10}} & \multirow{-3}{*}{\tablenum{1.50}} & \multirow{-3}{*}{\tablenum{4.00}} & \multirow{-3}{*}{\tablenum{1.50}} & \multirow{-3}{*}{\tablenum{0.50}} & \multirow{-3}{*}{\tablenum{-2.00}} & {$ \< N_\da \>$} & 2.62665(97) & 2.63011(69) & 2.6331(16) & 2.6328\\
\TableSectionSeperator
 &  &  &  &  &  &  &  & {$\< E \> $} & 21.8894(15) & 21.9257(15) & 21.9610(17) & 21.9587\\
 &  &  &  &  &  &  &  & {$ \< N_\ua \>$} & 7.67379(40) & 7.66349(41) & 7.65307(75) & 7.6524\\
\multirow{-3}{*}{\tablenum{3}} & \multirow{-3}{*}{\tablenum{2}} & \multirow{-3}{*}{\tablenum{10}} & \multirow{-3}{*}{\tablenum{4.80}} & \multirow{-3}{*}{\tablenum{5.00}} & \multirow{-3}{*}{\tablenum{1.50}} & \multirow{-3}{*}{\tablenum{0.50}} & \multirow{-3}{*}{\tablenum{2.00}} & {$ \< N_\da \>$} & 2.39772(48) & 2.41500(48) & 2.43218(63) & 2.4321\\
\TableSectionSeperator
 &  &  &  &  &  &  &  & {$\< E \> $} & 11.1153(22) & 11.1625(21) & 11.2111(30) & 11.2087\\
 &  &  &  &  &  &  &  & {$ \< N_\ua \>$} & 6.15373(26) & 6.15798(26) & 6.16227(40) & 6.1619\\
\multirow{-3}{*}{\tablenum{3}} & \multirow{-3}{*}{\tablenum{2}} & \multirow{-3}{*}{\tablenum{10}} & \multirow{-3}{*}{\tablenum{2.50}} & \multirow{-3}{*}{\tablenum{3.50}} & \multirow{-3}{*}{\tablenum{1.50}} & \multirow{-3}{*}{\tablenum{0.50}} & \multirow{-3}{*}{\tablenum{-2.00}} & {$ \< N_\da \>$} & 3.67981(43) & 3.69501(84) & 3.71093(45) & 3.7104\\
%-------------------------------------%
\BotRule
    \end{tabular}
  \caption{Comparison of the worm algorithm with exact results obtained by diagonalizing the Hamiltonian for small lattices in $d=1,2,3$ spatial dimensions.  To demonstrate that extrapolation to the $\varepsilon \to 0 $ limit is necessary for the agreement, we show an illustrative fit for the fourth row in \Cref{fig:epsilon-extrapolation-d3-U2}.
  \label{tab:exact-hamiltonian-comparison}}
\end{table*}

\section{Exact Results from Small Lattices}
\label{sec:exact-results-small-lattice}

In this appendix we compare our Monte Carlo results with exact results on small lattices over a wide range of parameters. This study helps verify the correctness of our algorithm and provides some benchmark calculations for readers to understand our model and verify our results if necessary.

\subsection{Discrete time results}

We first verify that our algorithm is able to reproduce the exact partition function on a small space-time lattice where we can enumerate all configurations. This should also help clarify the definition of the finite-$\varepsilon$ transfer matrix that we use. The exact expression for the finite-$\varepsilon$ 
partition function is given by (see \cref{zboson}) 
\begin{align}
    Z_\mu = \sum_{\MCConfig} \Omega(\MCConfig)
\end{align}
where $\MCConfig$ are the worldline configurations. On a $2 \times 2$ space-time lattice we can explicitly enumerate all configurations $\MCConfig$, which gives us
\begingroup
% \allowdisplaybreaks
\begin{align}
Z_\mu &= Z_0^\uparrow  Z_0^\downarrow  \\ 
&\quad + Z_0^\uparrow (2\ (W_m^{{\downarrow}})^2+8\ (W_m^{{\downarrow}})^2\, (W_h^{{\downarrow}})^2+(W_m^{{\downarrow}})^4) \nonumber\\
&\quad + Z_0^\downarrow (2\ (W_m^{{\uparrow}})^2+8\ (W_m^{{\uparrow}})^2\, (W_h^{{\uparrow}})^2+(W_m^{{\uparrow}})^4) \nonumber\\
&\quad +  2\ (W_m^{{\uparrow}})^2\, (W_m^{{\downarrow}})^2    \nonumber\\
&\quad +  2\ (W_m^{{\uparrow}})^2\, (W_m^{{\downarrow}})^2\, (W_I)^2    \nonumber\\
&\quad +  16\ (W_m^{{\uparrow}})^2\, (W_m^{{\downarrow}})^2\, (W_h^{{\uparrow}})^2\, W_I    \nonumber\\
&\quad +  16\ (W_m^{{\uparrow}})^2\, (W_m^{{\downarrow}})^2\, (W_h^{{\downarrow}})^2\, W_I    \nonumber\\
&\quad +  32\ (W_m^{{\uparrow}})^2\, (W_m^{{\downarrow}})^2\, (W_h^{{\uparrow}})^2\, (W_h^{{\downarrow}})^2    \nonumber\\
&\quad +  32\ (W_m^{{\uparrow}})^2\, (W_m^{{\downarrow}})^2\, (W_h^{{\uparrow}})^2\, (W_h^{{\downarrow}})^2\, (W_I)^2    \nonumber\\
&\quad +  2\ (W_m^{{\uparrow}})^2\, (W_m^{{\downarrow}})^4\, (W_I)^2    \nonumber\\
&\quad +  8\ (W_m^{{\uparrow}})^2\, (W_m^{{\downarrow}})^4\, (W_h^{{\uparrow}})^2\, (W_I)^2    \nonumber\\
&\quad +  2\ (W_m^{{\uparrow}})^4\, (W_m^{{\downarrow}})^2\, (W_I)^2    \nonumber\\
&\quad +  8\ (W_m^{{\uparrow}})^4\, (W_m^{{\downarrow}})^2\, (W_h^{{\downarrow}})^2\, (W_I)^2    \nonumber\\
&\quad +  (W_m^{{\uparrow}})^4\, (W_m^{{\downarrow}})^4\, (W_I)^4    \nonumber
\label{eq:exact-2x2-partition-function}
\end{align}
\endgroup
where $Z_0^\sigma = 1 + 4 (\WeightMove^\uparrow)^2 +4 (\WeightMove^\uparrow)^4$, and the local weights $\WeightMove^\sigma$, $\WeightHop^\sigma$, $\WeightInt$ were defined in \cref{sec:worldline-formulation} to be 
\begin{align*}
    W^{\sigma}_h &= \Hopping{\sigma}\varepsilon, \\
    W^{\sigma}_m &= \exp\big(-\varepsilon (2 d \Hopping{\sigma}- \mu_\sigma)\big),\\
    \WeightInt &= \exp\left(-\varepsilon U\right).
\end{align*}
Recall that the weight for each configuration is given by \cref{eq:weight-config},
so the number of particles $\NumParticles{\sigma}$, number of hops $\NumHops{\sigma}$ and number of interactions $\NumInteractions$ for each configuration can simply be read off from the exponents in expression \eqref{eq:exact-2x2-partition-function} above. The average energy is then computed using the definition \eqref{eq:energy-of-worldline}. \Cref{tab:exact-2x2} compares the results for the average particle numbers and average energy obtained from our Monte Carlo method with those from the exact expression in \cref{eq:exact-2x2-partition-function}.

\begin{figure*}
\centering
\includegraphics[width=\linewidth]{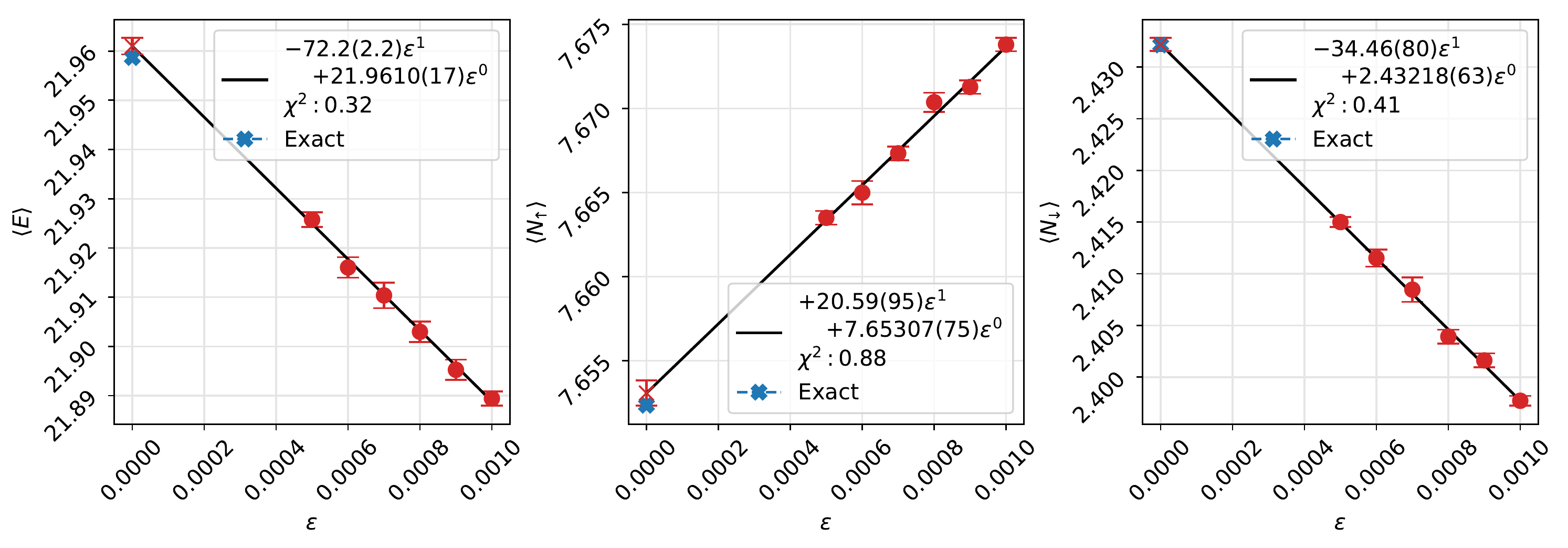}
\caption{Extrapolation 
to the continuous time limit $\varepsilon\to 0$ for $d = 3$, $\beta = 10$, $L_X = 2$, $\mu_\ua = 4.80$, $\mu_\da = 5.00$, $m_\ua = 1.50$, $m_\da = 0.50$, $U = 2.00$. This corresponds to the fourth row in \Cref{tab:exact-hamiltonian-comparison}.
}
\label{fig:epsilon-extrapolation-d3-U2}
\end{figure*}

\subsection{Continuous time results}

Since our time discretization procedure is very different from conventional approaches, it is useful to verify that our results do agree with the partition function \eqref{eq:Z-mu} in the $\varepsilon\to 0$ limit.  We can do this
for small spatial lattices by explicitly diagonalizing the Hamiltonian.
The dimension of the relevant Hilbert space for  $(\NN)$ hard-core bosons in a $d$-dimensional spatial lattice of size $\LX$ is
$\binom{\LX^d}{N_\uparrow} 
\binom{\LX^d}{N_\downarrow}$
where $N_\uparrow, N_\downarrow \in \{0, \dotsc, \LX^d \}$.

To compare with the exact Hamiltonian results, we take the  $\varepsilon \to 0$ limit by performing calculations at several values of $\varepsilon$  and performing a linear extrapolation. 
\cref{tab:exact-hamiltonian-comparison} shows a comparison of our Monte Carlo data with exact results at a few sets of parameters in $d=1,2,3$ dimensions for $\epsilon=\num{0.001}, \num{0.0005}$ and $\epsilon \to 0$.  We find that a proper extrapolation in higher dimensions is important to reproduce the exact results within errors.  We show an example of this in \cref{fig:epsilon-extrapolation-d3-U2}.  We believe this agreement provides another non-trivial check of our approach in arbitrary dimensions, coupling strengths, mass-imbalances and particle numbers.

\section{Second-Order Perturbation Theory}
\label{sec:perturbation-theory}

In many figures, we show results from first- and second-order perturbation theory.  Here we provide the expressions used to obtain these results. We assume the particles to be fermions for the Hamiltonian in \cref{eq:lattice-model}, and use 
$\EnergyGroundNNFermion$ to denote the
the ground-state energy for $(N_\ua, N_\da)$ fermions in a box of size $\LX$ (even) with periodic boundary conditions.  

The energy spectrum for the free theory ($U=0$) can be constructed using single particle energy eigenvalues which are characterized by integers $\vec k_\sigma = (k_{\sigma, 1}, \dotsc, k_{\sigma,\NumParticles{\sigma}})$ for $\sigma = \ua, \da$, which label the energy eigenstates of  individual fermions of each type.  Let the single $\sigma$-particle energy corresponding to the integer $k$ be  $\Hopping{\sigma}\epsilon(k)$ where
\begin{align}
\epsilon(k) &= 4 \sin^2 \left(\frac{\pi k}{\LX}\right)\quad \text{} k=0, \pm 1, \dotsc, \pm \LX/2.
\end{align}
The total energy of a system of $(\NumParticles{\uparrow}, \NumParticles{\downarrow})$ non-interacting fermions is then
\begin{align}
    \EnergyNNOrderZero(\vec k_\ua, \vec k_\da) = 
    \Hopping{\uparrow} \sum_{i=1}^{N_\uparrow} \epsilon({k_{\ua,i}}) + 
    \Hopping{\downarrow} \sum_{i=1}^{N_\downarrow} \epsilon({k_{\da, i}}).
\end{align}
When $\NumParticles{\sigma}$ is odd (the values which we consider in this paper) the ground state is unique and corresponds to the choice 
$\vec k_\sigma^0 \equiv \left(-{(\NumParticles{\sigma}-1)/2}, -{(\NumParticles{\sigma}-1)/2}+1, \dotsc, {(\NumParticles{\sigma}-1)/2}\right)$. The ground state energy of the full Hamiltonian in perturbation theory up to second order is then given by 
\begin{align}
\EnergyGroundNNFermion &= \EnergyGroundNNOrder{0}
    + U \EnergyGroundNNOrder{1} 
    + U^2 \EnergyGroundNNOrder{2} 
    + \Order(U^3)
\end{align}
where
\begin{align}
    \EnergyGroundNNOrder{0} &= \EnergyNNOrderZero(\vec k^0_\ua, \vec k^0_\da) \\
    \EnergyGroundNNOrder{1} &=  \frac{1}{\LX} N_\uparrow N_\downarrow \\
    \EnergyGroundNNOrder{2}  &=  
    \frac{1}{\LX^2}\ \sum_{ \mathclap{(\vec k_\ua, \vec k_\da)}  } {}^{'}
    \ \frac{1}{ \EnergyGroundNNOrder{0} - \EnergyNNOrderZero(\vec k_\ua, \vec k_\da) }
\end{align}
where the primed sum $\sum'$ is over states $(\vec k_\ua, \vec k_\da)$ that are related to the free ground state by an excitation of exactly one particle of each type, such that the total change in momentum is zero: $\vec k_\ua + \vec k_\da = \vec k^0_\ua + \vec k^0_\da$.

\bibliography{ref}

\end{document}